\documentclass[journal=jctcce,manuscript=article]{achemso}
\setkeys{acs}{usetitle=true}

\usepackage[version=3]{mhchem} 
\usepackage{longtable} 
\usepackage{makecell} 
\usepackage{amsmath,amssymb}
\usepackage{graphicx,color,subfigure}
\usepackage{amsfonts}
\usepackage{comment}
\usepackage{placeins}
\usepackage{fixltx2e}
\usepackage{braket} 

\usepackage{sectsty}
\usepackage{balance}
\usepackage{xspace}
\usepackage{algorithm,algorithmic}
\usepackage{appendix}
\usepackage{epsfig,amssymb,amsmath,amsthm,amsfonts,amsbsy,mathrsfs}
\usepackage{graphicx,color}
\usepackage{amsmath}
\usepackage{amssymb}
\usepackage{epstopdf}
\usepackage{threeparttable}
\usepackage{graphicx} 
\usepackage{lastpage}
\usepackage{indentfirst} 
\usepackage{bm}
\usepackage[format=plain,justification=raggedright,singlelinecheck=false,font=small,labelfont=bf,labelsep=space]{caption}

\allowdisplaybreaks

\usepackage[version=3]{mhchem} 
\usepackage{array}
\usepackage{booktabs}

\title{Perturbative Variational Quantum Eigensolver via Reduced Density Matrices}

\author{Yuhan Zheng}
\affiliation{Hefei National Research Center for Physical Sciences at the Microscale, University of Science and Technology of China, Hefei, Anhui 230026, China} 
\altaffiliation{These authors contributed equally.}

\author{Yibin Guo}
\affiliation{Beijing Academy of Quantum Information Sciences, Beijing, China}
\alsoaffiliation{Institute of Physics, Chinese Academy of Sciences, Beijing 100190, China}
\alsoaffiliation{University of Chinese Academy of Sciences, Beijing 101408, China}
\altaffiliation{These authors contributed equally.}

\author{Huili Zhang}
\affiliation{Beijing Key Laboratory of Fault-Tolerant Quantum Computing, Beijing Academy of Quantum Information Sciences, Beijing 100193, China}

\author{Jie Liu}
\email{liujie86@ustc.edu.cn} 
\affiliation{Hefei National Laboratory, University of Science and Technology of China, Hefei 230088, China}

\author{Xiongzhi Zeng}
\email{xzzeng@ustc.edu.cn} 
\affiliation{State Key Laboratory of Precision and Intelligent Chemistry, University of Science and Technology of China, Hefei, Anhui 230026, China} 

\author{Xiaoxia Cai}
\email{xxcai@ihep.ac.cn} 
\affiliation{Institute of High Energy Physics, Chinese Academy of Sciences, Beijing 100049, China}

\author{Zhenyu Li}
\email{zyli@ustc.edu.cn} 

\affiliation{State Key Laboratory of Precision and Intelligent Chemistry, University of Science and Technology of China, Hefei, Anhui 230026, China} 
\alsoaffiliation{Hefei National Laboratory, University of Science and Technology of China, Hefei 230088, China}

\author{Jinlong Yang}
\affiliation{State Key Laboratory of Precision and Intelligent Chemistry, University of Science and Technology of China, Hefei, Anhui 230026, China}
\alsoaffiliation{Hefei National Laboratory, University of Science and Technology of China, Hefei 230088, China}

\begin{document}

\begin{abstract}
Current noisy intermediate-scale quantum (NISQ) devices remain limited in their ability to perform accurate quantum chemistry simulations due to restricted numbers of high-fidelity qubits and short coherence times. To overcome these challenges, we introduce a reduced density matrix (RDM)-based perturbative variational quantum eigensolver (VQE) framework that augments active-space VQE with perturbation theory to recover electron correlation beyond the active space without increasing the qubit count or variational circuit depth. We formulate a fully coupled approach (VQE-PTs) and a diagonal approximation (VQE-PT). The former retains couplings among orthonormalized perturbers, whereas the latter neglects these couplings to simplify the classical post-processing. Numerical simulations of HF, N$_2$, and F$_2$ show that VQE-PTs provides a robust formulation across different molecular systems, while VQE-PT offers an efficient approximation. We further experimentally implement VQE-PT on the Quafu superconducting quantum processor for F$_2$, achieving a mean absolute error of 1.2 millihartree along the potential energy surface after error mitigation. These results demonstrate perturbative VQE as a practical framework for incorporating dynamic correlation in quantum chemistry simulations.
\end{abstract}

\section{Introduction}
An accurate treatment of electron correlation is fundamental to computational chemistry, as it underpins reliable predictions of molecular properties, reaction mechanisms, and the behavior of complex materials. Multireference perturbation theory (MRPT)~\cite{SzaMulGid12,LisNacAqu18} offers a balanced computational framework in which static correlation is treated by a (near-)exact active-space method, such as the complete active space self-consistent field (CASSCF)~\cite{ROOS1980157,Ols11,aquilante2016molcas}, selected configuration interaction~\cite{HolTubUmr16,LiuHof16,LevHaiTub20,TubFreLev20}, or the density matrix renormalization group~\cite{Whi92,ChaSha11,VerMalRoo11}, while dynamic correlation beyond the active space is incorporated via perturbation theory~\cite{pulay2011perspective,angeli2001introduction,townsend2019post}. By separating the treatment of static and dynamic correlation, this framework offers a favorable balance between accuracy and computational cost, and has become a standard approach for strongly correlated electronic structure problems.

The emergence of quantum computing provides an alternative route for solving active-space electronic structure problems with high accuracy~\cite{Fey82,CaoRomOls19,McAEndAsp20,LiuFanLi22}. Among the quantum algorithms proposed for this purpose, the variational quantum eigensolver (VQE)~\cite{PerMcCSha14,TilCheCao22} is one of the most popular approaches owing to its compatibility with near-term quantum hardware. As a hybrid quantum-classical algorithm, VQE employs a quantum processor to prepare trial states and evaluate expectation values, while a classical optimizer iteratively updates the wave function parameters. From the perspective of multireference theory, VQE can be viewed as a promising quantum counterpart to classical active-space solvers for describing static correlation. To recover dynamic correlation contributions beyond the active space, two complementary hybrid quantum-classical strategies have recently been explored. The first relies on Hamiltonian downfolding techniques~\cite{huang2023leveraging,bauman2026coupled}, in which external-space correlation effects are incorporated into an effective active-space Hamiltonian through a unitary transformation, allowing the subsequent quantum simulation to account for correlation contributions outside the active space. The second applies perturbative corrections to a quantum-computed active-space wave function. Along this line, the combination of variational quantum eigensolver (VQE) methods with perturbation theory (PT) has attracted considerable attention as a natural route to recovering dynamic correlation beyond the active space and achieving a more balanced description of electron correlation.~\cite{TamGalRic23,FitTalRem24,KroRam22,LieMotPel24}.

Despite this appealing conceptual framework, the practical implementation of perturbative VQE algorithms on near-term quantum devices still faces important  challenges. On the one hand, current noisy intermediate-scale quantum (NISQ) devices are constrained by finite coherence times and non-negligible gate errors~\cite{Gao2025Zuchongzhi3,GoogleWillow2024,JavMarHol25}. On the other hand, the perturbative correction must be formulated in a manner that is both accurate and compatible with the limited quantum resources available on near-term hardware. These considerations motivate the development of resource-efficient VQE-based perturbative methods that can recover correlation beyond the active space without substantially increasing circuit complexity. Recent advances along this direction include VQE-based formulations of second-order $N$-electron valence perturbation theory (NEVPT2)~\cite{TamGalRic23,FitTalRem24,KroRam22} and complete active space second-order perturbation theory (CASPT2)~\cite{LieMotPel24}. For instance, Liepuoniute {\it et al.} extracted a purified configuration interaction vector from quantum state tomography to serve as the reference for conventional CASPT2, and subsequently performed repeated classical calculations to average out the sampling uncertainty~\cite{LieMotPel24}. Fitzpatrick {\it et al.} demonstrated that the high-order RDMs necessary for NEVPT2 calculations can be reconstructed by reusing the measurement data from the initial ground-state preparation via adaptive informationally complete positive operator-valued measures~\cite{FitTalRem24}. Despite these advances, developing robust, computationally efficient, and hardware-friendly frameworks for integrating variational quantum algorithms with perturbation theory remains a key research objective.

In this work, we propose perturbative corrections to the VQE energy via reduced density matrices (RDMs)~\cite{coleman2007reduced}. Within this framework, VQE prepares a multireference wave function that captures static electron correlation within the active space, while perturbation theory recovers the remaining dynamic correlation beyond the active space. The perturbative correction is evaluated directly from RDMs measured on a quantum computer, thereby avoiding any increase in circuit width or depth. We benchmark several variants of perturbative VQE through numerical simulations and further experimentally demonstrate one scheme on the Quafu superconducting quantum processor by computing the \ce{F2} potential energy surface with an optimized quantum circuit. To suppress the impact of hardware noise, we employ two complementary post-processing strategies: (i) symmetry verification~\cite{PhysRevA.98.062339,sagastizabal2019experimental} to post-select the target symmetry sector, and (ii) $N$-representability-guided purification through semidefinite programming~\cite{coleman2007reduced,rubin2018application,cai2023quantum,smart2019quantum,mazziotti2012two}. With these techniques, the perturbative VQE attains chemical accuracy over most bond lengths and yields a mean absolute error of 1.2 millihartree (mHa) relative to the exact diagonalization energies.

\section{Theory}\label{sec:theory}

\begin{figure}[!htb]   
	\centering
\includegraphics[width=1.0\linewidth]{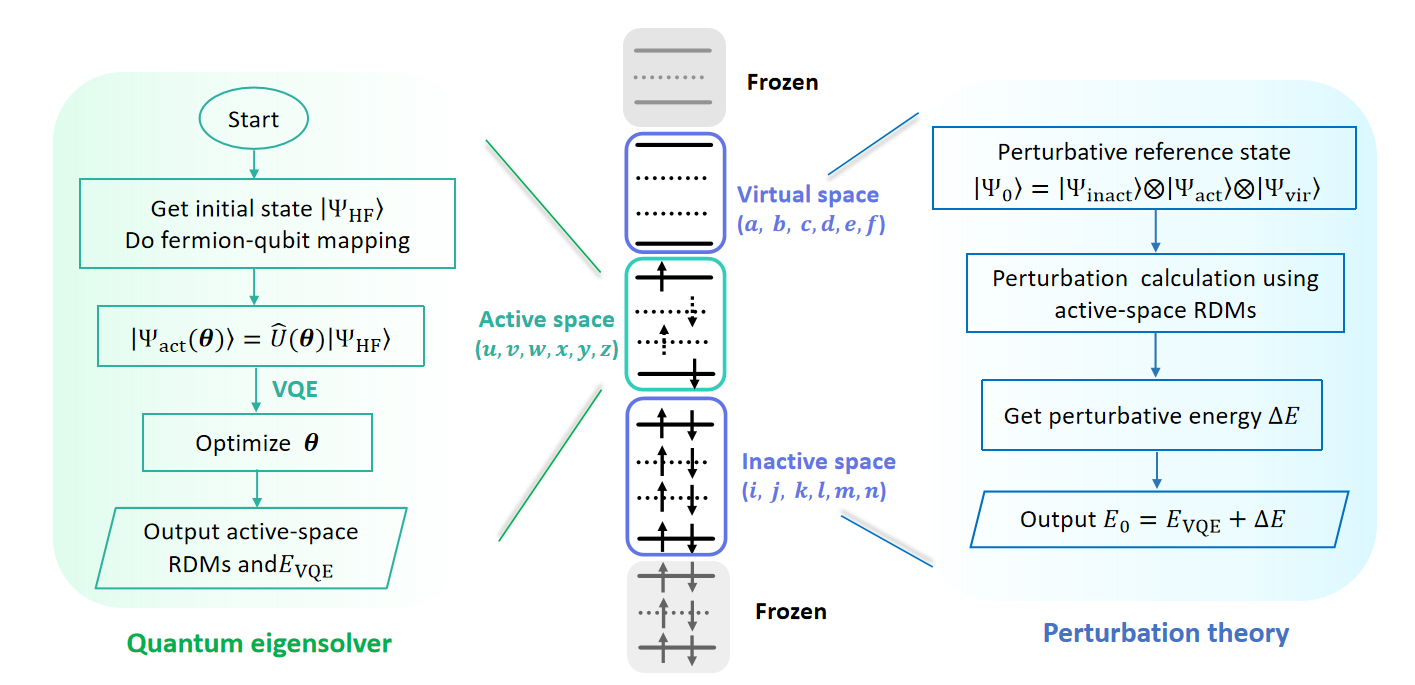}
	\caption{{Algorithm flowchart of the perturbative variational quantum eigensolver. The molecular orbital space is partitioned into active, perturbative and frozen spaces. The perturbative space is composed of the inactive and virtual spaces. The electron correlation in the active space is accounted for by the variational quantum eigensolver (VQE) calculations. The dynamical electron correlation beyond the active space is treated using perturbation theory.}}
	\label{fig:space}  
\end{figure}

Analogous to MRPT, the perturbative VQE method partitions the total orbital space into three regions: active, perturbative, and frozen (optional) spaces, as illustrated in Fig.~\ref{fig:space}. The perturbative space consists of the inactive and virtual spaces, which include doubly occupied inactive orbitals and unoccupied virtual orbitals, respectively. The perturbative VQE method employs VQE to compute the eigenvalue and corresponding wave function of the effective Hamiltonian defined in the active space, accounting for static electron correlation. The perturbative correction to the reference energy, accounting for dynamical correlation, is computed using the active-space wave function as the zeroth-order reference state.

\subsection{Variational Quantum Eigensolver}\label{subsec:VQE}
An active-space wave function is prepared by applying a sequence of unitary transformations to an initial state~\cite{GriEcoBar19,LiuLiYan21,burton2024accurate,endo2021hybrid}
\begin{equation}
    |\Psi_{\rm act}(\bm{\theta})\rangle = \hat{U}_M(\theta_M) \cdots \hat{U}_1(\theta_1) |\varphi_0\rangle. 
\end{equation}
Here, $\hat{U}_k$ is either a one-qubit or two-qubit gate operation. To solve an electronic structure problem, the initial state $|\varphi_0\rangle$ is chosen as the active-space component of the Hartree-Fock state. The chemically inspired form of unitary transformations $\hat{U}_\mu = e^{\theta_\mu \hat{\tau}_\mu}$ are often used to construct the wave function ansatz,~\cite{bauer2020quantum, anand2022quantum,VQE_1,VQE_2} with $\hat{\tau}_\mu= \hat{T}_\mu - \hat{T}^\dagger_\mu$ being an anti-Hermitian excitation operator. In this work, only the single- and double-excitation operators $\hat{T}_\mu \in \{\hat{a}^{\dagger}_u\hat{a}_v,\ \hat{a}^{\dagger}_u\hat{a}^{\dagger}_v\hat{a}_w\hat{a}_x\}$ are used in the VQE calculations. $u,\ v,\ w,\ x,\ \ldots$ denote spin orbitals in the active space. Inactive spin orbitals are labeled as $i,\ j,\ k,\ \ldots$, and virtual spin orbitals are labeled as $a,\ b,\ c\ \ldots$. The variational parameters are optimized through
\begin{equation}
    E_{\rm VQE} = \min_{\bm{\theta}} \langle \Psi_{\rm act}(\bm{\theta}) | \hat{H}_{\rm act} |\Psi_{\rm act}(\bm{\theta})\rangle,
\end{equation}
where $\hat{H}_{\rm act}$ is the effective Hamiltonian in the active space, namely 
\begin{equation}
\hat{H}_{\rm act} = \left( \bra{11..1}_{\rm inact} \otimes \bra{00..0}_{\rm vir} \right) \hat{H} \left( \ket{11..1}_{\rm inact} \otimes \ket{00..0}_{\rm vir} \right),
\end{equation}
where $\hat{H}$ is the electronic Hamiltonian. For brevity in the subsequent discussion, we denote the inactive state $ \ket{11..1}_{\rm inact}$ as $\ket{\Psi_{\rm inact}}$ and the virtual state $ \ket{00..0}_{\rm vir}$ as $\ket{\Psi_{\rm vir}}$. Adaptive derivative-assembled pseudo-Trotter (ADAPT) VQE~\cite{grimsley2019adaptive} is used to iteratively construct the ansatz in the subsequent numerical simulations, and the optimized parameters are denoted as $\boldsymbol{\theta}_0$.
 
 \subsection{Perturbation Theory}\label{sec:PT}
 The reference wave function of the ground state is defined as the tensor product of the inactive, active, and virtual spatial components:
\begin{equation}\label{eq:psi_0}
\begin{split}
    | \Psi_0 \rangle = |\Psi_{\rm inact}\rangle\otimes|\Psi_{\rm act}(\boldsymbol{\theta}_0)\rangle\otimes|\Psi_{\rm vir}\rangle
\end{split}
\end{equation}
 To recover the dynamic correlation energy missing from the active-space simulation, the first-order wave function $| \Psi^{(1)} \rangle$ lies in the linear space spanned by a set of $K$ internally contracted perturbers, denoted as $\{|\Phi_I\rangle\}_{I=1}^K$~\cite{werner1988efficient}. These perturbers are generated by applying spin-adapted excitation operators $\hat{r}_I$ directly to the reference state, i.e., $|\Phi_I\rangle=\hat{r}_I|\Psi_0\rangle$~\cite{helgaker2013molecular, chan2021molecular}. 


Since the ADAPT-VQE optimization is terminated when the operator-gradient norm falls below the prescribed threshold, the Hamiltonian matrix elements coupling the reference state to excitations that are purely internal to the active space are numerically negligible:
 \begin{equation}\label{eq:couping}
 \langle \Psi_{\rm act} | \hat{H} \hat{r}_{uv} | \Psi_{\rm act} \rangle \approx 0, 
 \qquad
 \langle \Psi_{\rm act} | \hat{H} \hat{r}_{uvwx} | \Psi_{\rm act} \rangle \approx 0,
 \end{equation}
 for active indices $u,v,w,x$. Numerical verification of Eq.~\eqref{eq:couping} for three molecular systems considered in this work is provided in Supporting Information. Therefore, the perturbative operator pool $\{\hat{r}_I\}$ is restricted to excitations that involve at least one orbital outside the active space (inactive or virtual). 

In the internally contracted approach, the perturbers $\{\ket{\Phi_I}\}$ are generally nonorthogonal, and the corresponding overlap matrix is block diagonal with respect to the excitation manifolds. For perturbers belonging to the $\alpha$-th excitation manifold $\mathcal{M}_\alpha$, the overlap matrix is defined as
\[
S_{IJ}^{(\alpha)} = \braket{\Phi_I | \Phi_J}, \qquad I,J \in \mathcal{M}_\alpha,
\]
which generally satisfies $S_{IJ}^{(\alpha)} \neq \delta_{IJ}$. To formulate the perturbative VQE framework within standard perturbation theory~\cite{sakurai2020modern}, we construct an orthonormal basis for each excitation manifold using block-wise symmetric (L\"owdin) orthogonalization:
\begin{equation}
\label{eq:lowdin}
\ket{\Psi_J^{(\alpha)}} =
\sum_{I \in \mathcal{M}_\alpha}
\left(\mathbf{S}^{(\alpha)}\right)^{-1/2}_{IJ}
\ket{\Phi_I}.
\end{equation}
For notational simplicity, the manifold label $\alpha$ is omitted in the following discussion, and the orthonormal perturber states are collectively denoted by $\{\ket{\Psi_I}\}_{I=1}^{K'}$. The total dimension of the resulting orthonormal perturbative subspace is denoted by $K'$, where $K' \le K$.

With the orthonormal basis $\{\ket{\Psi_0}, \ket{\Psi_1}, \dots, \ket{\Psi_{K'}}\}$, we formulate perturbation theory by partitioning the full Hamiltonian as $\hat{H} = \hat{H}_0 + \hat{V}$. The zeroth-order Hamiltonian is defined as the projection of $\hat{H}$ onto the reference state and the orthonormal perturbative subspace:
\begin{equation}
\label{eq:H0_coupled}
\hat{H}_0
=
E_{\rm VQE}\ket{\Psi_0}\bra{\Psi_0}
+
\sum_{I,J=1}^{K'}
\tilde{H}_{IJ}\ket{\Psi_I}\bra{\Psi_J},
\end{equation}
where
\[
E_{\rm VQE}=\bra{\Psi_0}\hat{H}\ket{\Psi_0}, \qquad
\tilde{H}_{IJ}=\bra{\Psi_I}\hat{H}\ket{\Psi_J},
\]
and $\tilde{H}_{IJ}$ are the Hamiltonian matrix elements evaluated within the orthonormal perturber basis. The perturbation operator is then defined as $\hat{V}=\hat{H}-\hat{H}_0$.

By construction, the reference state $\ket{\Psi_0}$ is an eigenstate of $\hat{H}_0$ with eigenvalue $E_{\rm VQE}$, and the first-order energy correction vanishes ($E^{(1)} = \bra{\Psi_0}\hat{V}\ket{\Psi_0} = 0$). The first-order correction to the wave function, $\ket{\Psi^{(1)}}$, therefore satisfies
\begin{equation}\label{eq:RSPT_1}
\left(\hat{H}_0 - E_{\rm VQE}\right)\ket{\Psi^{(1)}} = -\hat{V}\ket{\Psi_0}.
\end{equation}
Expanding $\ket{\Psi^{(1)}}$ in the orthonormal perturber basis,
\[
\ket{\Psi^{(1)}}=\sum_{J=1}^{K'} \tilde{c}_J \ket{\Psi_J},
\]
and projecting onto $\bra{\Psi_I}$ gives
\begin{equation}
\sum_{J=1}^{K'} \left(\tilde{H}_{IJ}-E_{\rm VQE}\delta_{IJ}\right)\tilde{c}_J = \tilde{V}_I,
\end{equation}
where
\[
\tilde{V}_I = -\bra{\Psi_I}\hat{V}\ket{\Psi_0}
           = -\bra{\Psi_I}\hat{H}\ket{\Psi_0}.
\]
Solving this linear system yields the first-order amplitudes $\{\tilde{c}_I\}$, and the perturbative energy correction is then given by
\begin{equation}
\Delta E_{\mathrm{VQE-PTs}} = \sum_{I=1}^{K'} \tilde{c}_I \bra{\Psi_0}\hat{H}\ket{\Psi_I}.
\end{equation}
We denote this fully coupled approach as VQE-PTs. 

To avoid solving the coupled first-order amplitude equations and reduce the associated classical post-processing overhead, we also consider a diagonal partitioning scheme. In this approximation, the zeroth-order Hamiltonian retains only the diagonal matrix elements in the orthonormal perturber basis
\begin{equation}\label{eq:H0_diagonal}
    \hat{H}^{\rm D}_0 = E_{\rm VQE} |\Psi_0\rangle\langle\Psi_0| + \sum_{I=1}^{K'} \tilde{H}_{II} |\Psi_I\rangle\langle\Psi_I|.
\end{equation}
Under this approximation, Eq.~\eqref{eq:RSPT_1} is fully decoupled, so that the coefficients can be written analytically as
\[
\tilde{c}_I=\frac{\tilde{V}_I}{\tilde{H}_{II}-E_{\rm VQE}}.
\]
The corresponding perturbative energy correction is then given by
\begin{equation}
\label{eq:PT2}
\Delta E_{\rm VQE\text{-}PT}
=
\sum_{I=1}^{K'}
\frac{\left|\bra{\Psi_I}\hat{H}\ket{\Psi_0}\right|^2}
{E_{\rm VQE}-\tilde{H}_{II}}.
\end{equation}
We refer to this decoupled, diagonal approach as VQE-PT. Throughout this work, we investigate two variants of the perturbative VQE framework. The fully coupled formulation (VQE-PTs) provides the general framework, while the diagonal approximation (VQE-PT) offers a classically simplified implementation. Within the present RDM-based implementation, both variants can be evaluated from the same quantum-measurement data: both require the overlap and Hamiltonian matrices in the nonorthogonal perturber basis, which are constructed from active-space RDMs up to the same maximum rank.


\section{Numerical Results} 

\subsection{Performance of Perturbative VQE}\label{HF_and_N2}

We first assess the performance of  VQE-PTs and VQE-PT through noiseless numerical simulations of the ground-state dissociation curves for three molecules: hydrogen fluoride (\ce{HF}), nitrogen molecule (\ce{N2}), and fluorine molecule (\ce{F2}). We compare the results of VQE-PTs and VQE-PT with the standard VQE (with the same active space) and the full configuration interaction (FCI) reference values. These calculations were carried out using the quantum computational chemistry software $\rm Q^2$Chemistry~\cite{fan2022q}. The Jordan--Wigner transformation~\cite{JorWig28, fradkin1989jordan} is employed for fermion-to-qubit mapping. The STO-3G basis set is used for \ce{HF} and \ce{N2}, whereas the STO-6G basis set is adopted for \ce{F2}.

For \ce{HF} in the $C_{2v}$ point group, the two highest-energy $A_1$ orbitals are chosen as the active space, yielding a (2e, 2o) active space, while all remaining orbitals are included in the perturbative space. For \ce{N2} in the $D_{2h}$ point group, the active space comprises six molecular orbitals, namely $B_{2u}$, $B_{3u}$, $A_g$, $B_{2g}$, $B_{3g}$, and $B_{1u}$, corresponding to a (6e, 6o) active space. All remaining orbitals are assigned to the perturbative space and treated at the perturbation-theory level. For \ce{F2}, a minimal (2e, 2o) active space is employed, consisting of the $\sigma_g$ and $\sigma_u^*$ orbitals, while the two lowest-energy $1s$ core orbitals are kept frozen. In this case, all remaining non-frozen orbitals outside the active space are treated as the perturbative space. 

\begin{figure}[H]   
	\centering	\includegraphics[width=1\linewidth]{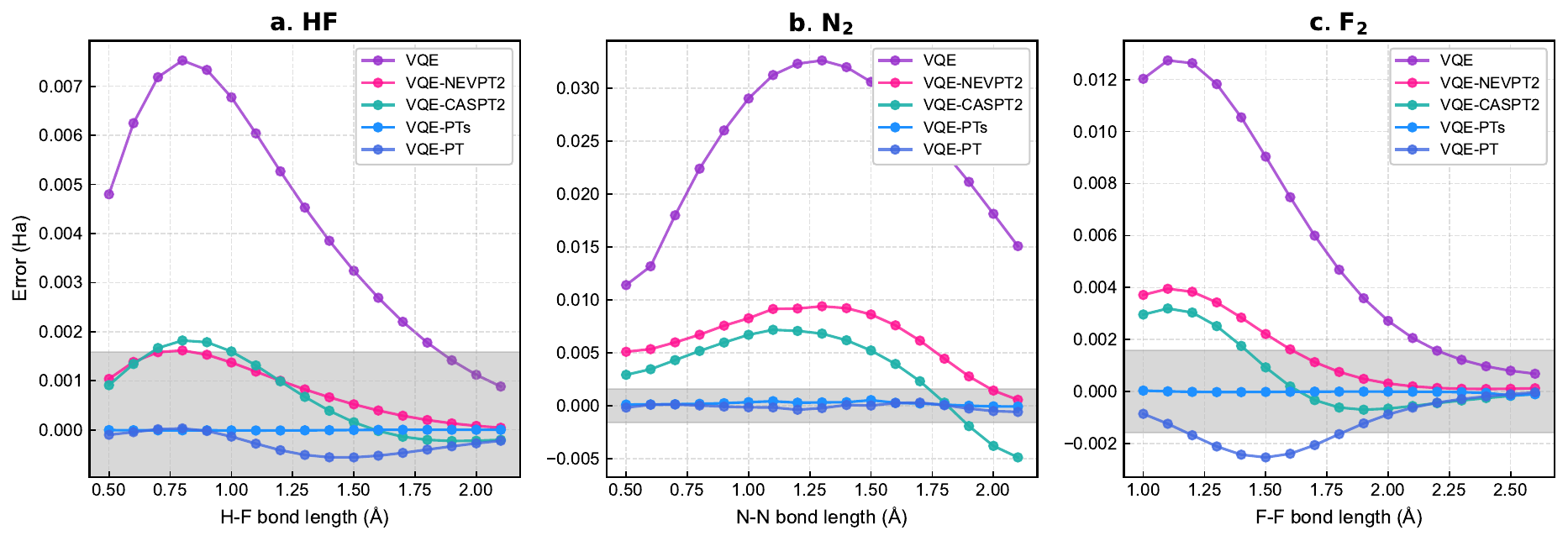}
	\caption{{Energy deviations (in Ha) of active-space VQE, VQE-PT, VQE-PTs, VQE-NEVPT2, and VQE-CASPT2, with respect to the full configuration interaction (FCI) reference energies. Results are shown for (a) HF and (b) N$_2$ in the STO-3G basis, and for (c) F$_2$ in the STO-6G basis. 
    The shaded region indicates the chemical accuracy threshold ($\approx 1.6$ mHa).}}
	\label{fig:vs3PT}
\end{figure}

Figure~\ref{fig:vs3PT} compares the energy deviations of all evaluated methods with respect to the FCI reference. The active-space VQE exhibits deviations on the order of $10^{-3}$--$10^{-2}$ Ha, reflecting the missing dynamic correlation associated with excitations outside the active space. Introducing perturbative corrections substantially reduces these errors without increasing the active-space size or quantum circuit complexity. For HF and N$_2$, VQE-PT closely follows the more rigorous VQE-PTs results and remains within chemical accuracy over most of the dissociation curves, indicating that the diagonal approximation captures the dominant perturbative contributions in these systems. In contrast, for F$_2$, VQE-PTs remains close to the FCI reference, whereas VQE-PT shows noticeable overcorrection around $1.5$ Å, suggesting that stronger couplings between active and external spaces make the diagonal approximation less reliable. Nevertheless, VQE-PT still provides substantial improvement over bare VQE.
 
\subsection{Comparison with Multireference Perturbation Theory}\label{three_PT}

We next evaluate the performance of both variants of our perturbative VQE framework against two MRPT schemes: CASPT2~\cite{andersson1992second} and NEVPT2~\cite{angeli2002n, angeli2001n}. To clarify the relationship between the present approach and existing internally contracted MRPT methods, we outline the fundamental theoretical distinctions among these methods. These distinctions primarily arise from two aspects: the construction of the zeroth-order Hamiltonian ($\hat H_0$) and the representation of the external perturber space.

Regarding the definition of $\hat{H}_0$, the CASPT2 zeroth-order Hamiltonian is constructed from a generalized one-body Fock operator $\hat{F}$ with appropriate projections onto the reference and external spaces. Although computationally convenient, this partitioning can place some external perturbers close in energy to the reference state, giving rise to small energy denominators and the well-known intruder-state problem, which is commonly mitigated using level-shift techniques.~\cite{roos1995multiconfigurational} NEVPT2 instead employs the Dyall Hamiltonian $\hat{H}^{D}$~\cite{dyall1995choice}, which explicitly retains the two-electron interactions within the active space and renders standard single-state NEVPT2 intrinsically free from the conventional intruder-state problem.~\cite{sokolov2024multireference} In contrast to both approaches, the perturbative VQE framework constructs $\hat{H}_0$ directly from the projection of the full electronic Hamiltonian $\hat{H}$ onto the orthonormalized perturbative subspace. VQE-PTs retains the complete projected Hamiltonian (Eq.~\ref{eq:H0_coupled}), whereas VQE-PT retains only its diagonal elements (Eq.~\ref{eq:H0_diagonal}). In both cases, two-electron interaction contributions are therefore naturally incorporated into $\hat{H}_0$. No intruder-state behavior was observed for the molecular systems considered here, although whether either formulation is intrinsically free from intruder states remains to be established.
 
 The second major distinction lies in the representation and orthogonalization of the external perturber basis. Standard CASPT2 utilizes a non-orthogonal internally contracted basis. Consequently, obtaining the perturbation amplitudes necessitates solving a large set of coupled linear equations. Alternatively, the partially contracted (PC) scheme of NEVPT2 retains the internally contracted perturbers within specific excitation classes (the $S_l^{(k)}$ manifolds) and diagonalizes the Dyall Hamiltonian within these subspaces to generate strictly orthogonal perturber states. Our perturbative VQE framework instead applies direct L\"owdin orthogonalization (Eq.~\ref{eq:lowdin}) to the spin-adapted internally contracted perturbers. The fully coupled VQE-PTs formulation retains the resulting off-diagonal Hamiltonian couplings and solves the coupled amplitude equations, whereas the diagonal VQE-PT approximation decouples these equations (Eq.~\ref{eq:PT2}) and bypasses the solution of coupled linear equations.

To assess the influence of different zeroth-order Hamiltonians and representations of the external perturber space, we employ VQE reference states as inputs to conventional PC-NEVPT2 and CASPT2 calculations, enabling a direct comparison with the perturbative VQE framework. The ADAPT-VQE algorithm is used to generate highly accurate reference states, whose energies and wavefunctions closely approximate the corresponding complete active space configuration interaction (CASCI) solutions. (With orbital optimization, this VQE baseline can be further extended toward the CASSCF level~\cite{sokolov2020quantum}.) VQE-NEVPT2 calculations were performed using the Prism library~\cite{prism_github} interfaced with PySCF~\cite{sun2020recent}, while VQE-CASPT2 was implemented in-house and validated against OpenMolcas~\cite{li2023openmolcas}. The HF, N$_2$, and F$_2$ molecules were studied using the same active and perturbative spaces as described in Section~\ref{HF_and_N2}.

In Fig.~\ref{fig:vs3PT}, all perturbative corrections, including VQE-PTs, VQE-PT, VQE-NEVPT2, and VQE-CASPT2, substantially improve upon bare VQE, but their performance differs along the dissociation coordinate. For HF and N$_2$, VQE-PT closely tracks VQE-PTs and generally yields smaller errors than VQE-CASPT2 and VQE-NEVPT2. For F$_2$, VQE-PTs remains close to the FCI reference, whereas VQE-PT overcorrects in part of the dissociation region; VQE-CASPT2 and VQE-NEVPT2 provide intermediate improvements. Taken together, these results identify VQE-PTs as the more robust accuracy-oriented formulation and VQE-PT as a useful classically simplified approximation whose accuracy depends on the strength of couplings within the perturbative subspace. Further tests with an enlarged 6-31G basis set for N$_2$ (see the Supporting Information) confirm this conclusion, demonstrating that VQE-PTs remains robust ($\sim$1 mHa error) and outperforms other perturbative variants when treating larger external spaces.

A key advantage of the perturbative VQE framework is that incorporating the perturbative correction does not require additional qubits or an enlarged variational circuit, since the missing dynamic correlation is recovered through post-processing of the active-space wave function. This advantage, however, comes at the cost of additional measurement overhead, as all four perturbative variants considered here ultimately require active-space RDMs up to the 4-RDM level (see the Supporting Information for details). The number of 4-RDM elements scales as $\mathcal{O}(N_{\rm act}^8)$ with the number of active spin orbitals, leading to a rapidly increasing measurement burden as the active space is enlarged. Although this overhead remains confined to the active space and is independent of the size of the external orbital space, it can become a significant scalability bottleneck for larger active spaces. Consequently, the present implementation is most practical for small- to medium-sized active spaces.

\section{Experimental Results}\label{exp}

\subsubsection{Experimental Setup}\label{exp_setup}
\begin{figure}[!htb]   
	\centering	\includegraphics[width=1\linewidth]{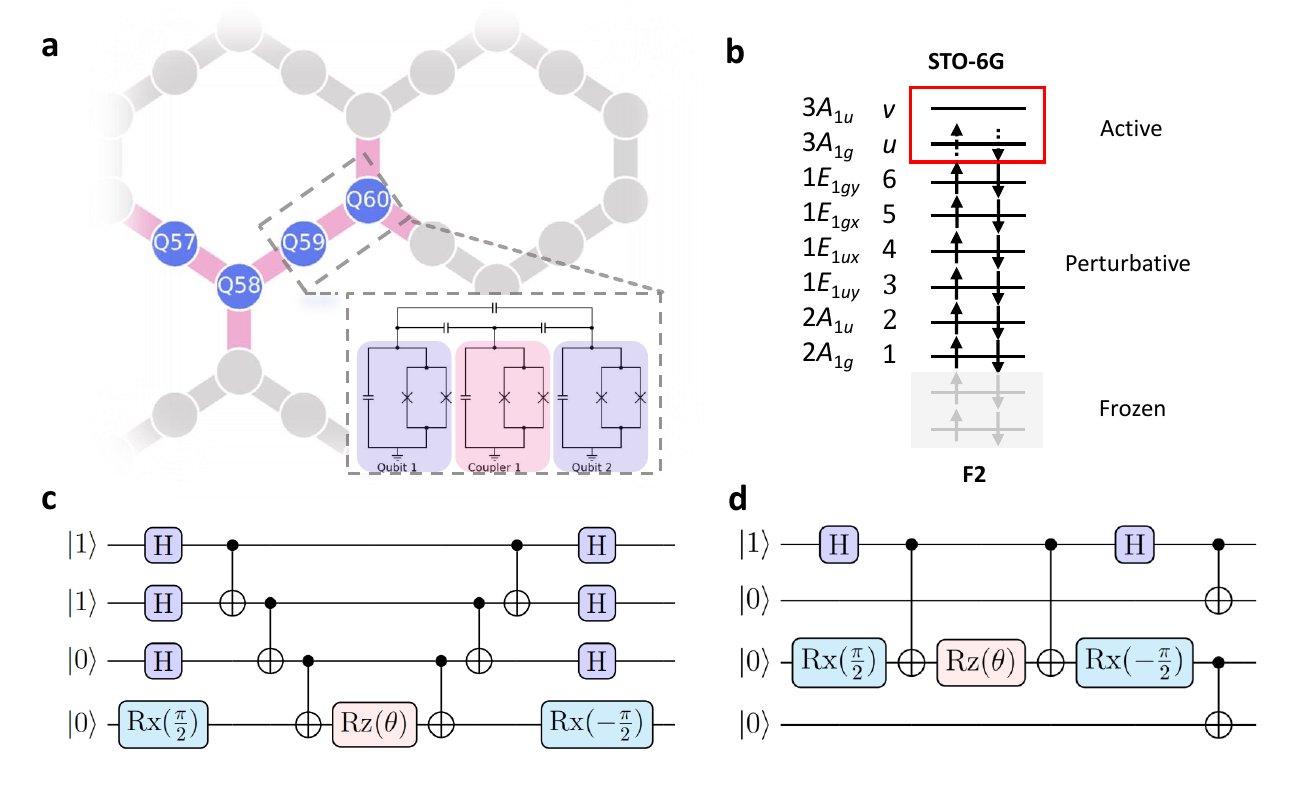}
	\caption{{(a) Connectivity topology of the Baihua processor; utilized qubits are highlighted. (b) Orbital partitioning for $\rm{F}_2$ (STO-6G) into frozen core, active space (2e, 2o), and perturbative space. The orbital symmetries are labeled according to the $D_{2h}$ point group, which is the maximal Abelian subgroup of $D_{\infty h}$ employed in the quantum chemistry calculations. (c) The circuit compilation of $e^{-i\frac{\theta}{2} X_0 X_1 X_2 Y_3}$. (d) Optimized ansatz circuit used in experiments, reducing the number of two-qubit gates from 6 to 4.}}
	\label{fig:circuit}
\end{figure}

We applied the VQE-PT method to compute the ground-state energy of $\rm{F}_2$ on the Quafu quantum cloud platform~\cite{Label}, utilizing the Baihua superconducting processor. As illustrated in Fig.~\ref{fig:circuit}(a), four linearly connected qubits (Q57 - Q60) were selected based on calibration data to execute the quantum circuits (see Supporting Information for detailed device parameters).

The electronic Hamiltonian was constructed using the STO-6G basis set with the Jordan--Wigner mapping.~\cite{JorWig28, fradkin1989jordan}. For the $\rm{F}_2$ molecule ($D_{\infty h}$ symmetry), we selected a minimal active space consisting of the frontier HOMO ($\sigma_g$) and LUMO ($\sigma_u^{*}$) bonding-antibonding orbital pair (2e, 2o), which captures the dominant static correlation along the dissociation coordinate. Two lowest-energy core $1s$ orbitals were frozen (Fig.~\ref{fig:circuit}(b)), and the remaining valence orbitals were treated within the perturbative space. While larger active spaces are desirable for quantitative accuracy, this (2e, 2o) configuration serves as an ideal testbed for two reasons. First, it captures the leading static correlation associated with bond stretching but misses a significant portion of dynamic correlation (see detailed validation in Supporting Information). This deficiency provides a benchmark to explicitly test the capacity of the VQE-PT framework to recover the missing energy. Second, it enables the construction of a shallow, hardware-efficient ansatz compatible with the coherence time and gate fidelity of NISQ devices, ensuring that the results reflect algorithmic performance rather than being dominated by hardware noise.

Within this active space, the VQE ansatz was chosen as
\begin{equation}
\hat{U}(\theta)\ket{\Psi_{\rm HF}} = e^{-i\frac{\theta}{2} X_0 X_1 X_2 Y_3}\ket{1100},
\end{equation}
which prepares a superposition of the two dominant configurations $\ket{1100}$ and $\ket{0011}$ (Fig.~\ref{fig:circuit}(c)). Because the target ground state belongs to the $A_g$ irreducible representation, pair-breaking configurations involving one electron in each active orbital transform as $A_u$ ($A_g \otimes A_u = A_u$) and are therefore excluded by spatial symmetry. Consequently, this problem can be reduced to an effective single-qubit representation. In the present work, however, we retain the full four-qubit representation of the active space rather than performing this additional symmetry reduction. This choice provides a more general demonstration of the VQE-PT framework and avoids relying on problem-specific reductions that are not necessarily available for larger or less symmetric systems. 

To reduce hardware-induced errors, the circuit was further optimized by decreasing the number of two-qubit gates from six to four, yielding an equivalent final state, shown in Fig.~\ref{fig:circuit}(d). The optimized ansatz is
\begin{equation}
\hat{U}'(\theta)\ket{\Psi_{\rm HF}} = [e^{-i\frac{\theta }{2} X_0 Y_2}{\rm CNOT}(01){\rm CNOT}(23)]\ket{1000}.
\end{equation}

After the VQE procedure, the perturbative reference wave function was defined as $| \Psi_0 \rangle = |\rm \Psi_{\rm inact}\rangle\otimes|\Psi_{\rm act}\rangle$, where $|\rm \Psi_{\rm inact}\rangle$ is the state with all perturbative orbitals doubly occupied.
We considered all symmetry-allowed excitations that promote electrons from the doubly occupied perturbative space to the active space ($\{\hat{r}_{2}^{v},\ \allowbreak  \hat{r}_{u 1}^{v \bar v},\  \allowbreak\hat{r}_{1\bar 1}^{v \bar v},\ \allowbreak \hat{r}^{v \bar v}_{2\bar 2},\ \allowbreak \hat{r}^{v \bar v}_{3\bar 3},\ \allowbreak \hat{r}^{v \bar v}_{4\bar 4},\ \allowbreak \hat{r}^{v \bar v}_{5\bar 5},\ \allowbreak \hat{r}^{v \bar v}_{6\bar 6} \}$). Crucially, because the (2e, 2o) active space contains only two electrons, all active-space RDMs of order higher than two are strictly zero. Consequently, evaluating the exact perturbative energy correction for this specific system requires measuring only up to the 2-RDM, completely avoiding the high measurement overhead typically associated with higher-order RDMs.

To evaluate the energy correction in the VQE-PT method, a total of 22 relevant Pauli strings were required. By exploiting qubit-wise commutativity, these observables were efficiently grouped into 9 distinct measurement bases to minimize hardware sampling overhead (see Supporting Information for details). Experiments were repeated five times for each data point with 1024 shots per Pauli string measurement.  After readout error mitigation~\cite{van2022model}, all experimental data were further processed using symmetry verification (SV)~\cite{PhysRevA.98.062339,sagastizabal2019experimental} and $N$-representability-based 2-RDM reconstruction~\cite{coleman2007reduced, rubin2018application,cai2023quantum,smart2019quantum,mazziotti2012two}, as described below.

\subsection{Post-Processing Techniques}
To perform electronic structure calculations on noisy quantum devices, it is necessary to utilize post-processing techniques to suppress the influence of noise~\cite{cai2023quantum, tilly2022variational}. In this work, we employ SV and $N$-representability-based techniques to improve the accuracy of calculations.

\subsubsection{Symmetry Verification}
If the Hamiltonian $\hat{H}$ commutes with the symmetry operator $\hat{S}$, i.e., 
\begin{equation}
\hat{H}\hat{S}=\hat{S}\hat{H},
\end{equation}
nondegenerate ground and excited states must reside in one of the eigenspaces of $\hat{S}$ ($\hat{S}\ket{\Psi}=s\ket{\Psi}$). However, during quantum computation, noise may cause the prepared state to acquire components outside the target eigenspace. SV~\cite{PhysRevA.98.062339,sagastizabal2019experimental} can be implemented either in-circuit, namely constraining the quantum circuit to preserve the target symmetry during execution, or via post-selection. In both cases, SV mitigates noise by ensuring that the ground or excited states remain within the desired subspace, enforced by the projector $\hat{M}_s=\frac{1}{2}(1+s\hat{S})$. The expectation value of $\hat{P}$ after SV is
\begin{equation}\label{sv}
\begin{split}
{\rm Tr}(\hat{P}\rho_s)&={\rm Tr}\left[\hat{P}\frac{\hat{M}_s\rho\hat{M}_s}{{\rm Tr}(\hat{M}_s\rho)}\right]\\
&=\frac{{\rm Tr}(\hat{P}\rho)+s{\rm Tr}(\hat{P}\hat{S}\rho)}{1+s{\rm Tr}(\hat{S}\rho)},
\end{split}
\end{equation}
where $\rho$ denotes the calculated density matrix and $\rho_s$ is the density matrix after SV.

In the present experiment, the effective Hamiltonian in the active space commutes with the symmetry operator $\hat{S}=Z_0Z_1Z_2Z_3$, and the ground state belongs to the $s=+1$ eigensubspace. We applied SV via classical post-processing exactly following Eq.~\eqref{sv}, where $\hat{P}$ iterates over all 22 Pauli strings required for the energy correction. By combining the expectation values of $\hat{P}$, $\hat{P}\hat{S}$, and $\hat{S}$ obtained after readout error mitigation, observables were reconstructed to strictly enforce the $s=+1$ symmetry.

\subsubsection{$N$ Representability}
The $N$-representability conditions impose constraints to ensure that a reconstructed $k$-RDM corresponds to a physically valid $N$-particle system.~\cite{coleman2007reduced, rubin2018application,cai2023quantum,smart2019quantum,mazziotti2012two} To reconstruct the 2-RDM from the symmetry-verified state $\rho_s$, we define the two-particle ($^2D$), two-hole ($^2Q$), and particle-hole ($^2G$) RDMs as
\begin{equation}
\begin{split}
^2D_{rs}^{pq}&={\rm Tr}(\hat{a}_p^\dagger\hat{a}_q^\dagger\hat{a}_s\hat{a}_r\, \rho_s),\\
^2Q_{rs}^{pq}&={\rm Tr}(\hat{a}_p\hat{a}_q\hat{a}_s^\dagger\hat{a}_r^\dagger\, \rho_s),\\
^2G_{rs}^{pq}&={\rm Tr}(\hat{a}_p^\dagger\hat{a}_q\hat{a}_s^\dagger\hat{a}_r\, \rho_s).
\end{split}
\end{equation}
The $2$-representability conditions require that these matrices to be positive semidefinite, referred to as the 2-positivity conditions:
\begin{equation}
\begin{split}
^2D\succeq 0,\quad
^2Q\succeq 0,\quad
^2G\succeq 0
\label{eq:SD}
\end{split}
\end{equation}
To reconstruct a physically valid 2-RDM, we solve the following semidefinite programming (SDP) problem~\cite{coleman2007reduced,mazziotti2011large,smart2019quantum,SmaBoyMaz22}:
\begin{equation}
\begin{array}{rl}
    \min  & E(\mathbf{x}) = \mathbf{c}^T \mathbf{x}  \\
    \text{s.t.} & \mathbf{A}\mathbf{x} = \mathbf{b}, \quad \mathbf{M}(\mathbf{x}) \succeq 0,
\end{array}
\end{equation}
where the column vector $\mathbf{c}$ encodes the one- and two-electron integrals of the Hamiltonian, and $\mathbf{x}$ contains all elements of the relevant RDMs. The equality $\mathbf{Ax=b}$ represents linear constraints among these RDMs (such as trace conditions and mapping between $^2D, ^2Q,$ and $^2G$), while the block-diagonal matrix constraint $\mathbf{M(x)} \succeq 0$ enforces the positivity conditions.

In this work, the experimentally measured Pauli expectation values are first converted into the corresponding RDM elements. These RDMs provide the quantum-derived initial point for the SDP procedure. The SDP procedure acts as a rigorous physical filter to strip away non-physical hardware noise and determines the final $N$-representable RDMs. This additional classical post-processing cost follows the same asymptotic scaling as conventional variational 2-RDM optimization. The number of independent 2-RDM variables scales as \(\mathcal{O}(N_{\rm act}^4)\) with the number of active spin orbitals, while the dominant matrix diagonalization steps in the SDP iterations scale approximately as \(\mathcal{O}(N_{\rm act}^6)\).

\subsection{Potential Energy Surface of $\bf{F_2}$}\label{result}
   
\begin{figure}[!htb]   
	\centering	\includegraphics[width=0.75\linewidth]{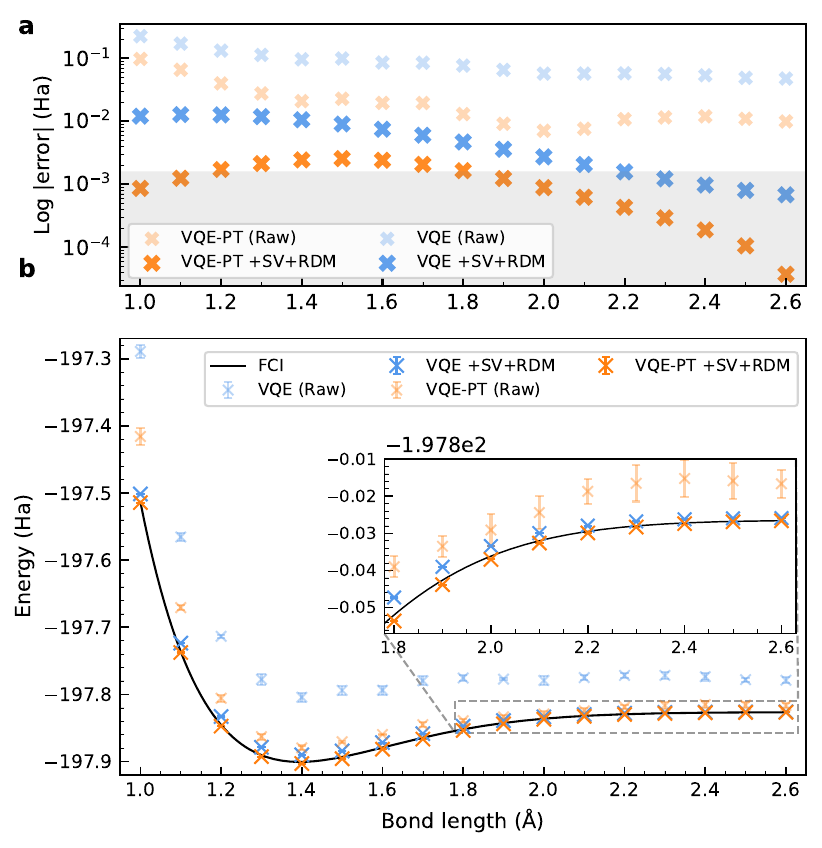}
	\caption{{Experimental ground-state energies (in Ha) of the F$_2$ molecule.
(a) Log-scaled energy errors of VQE and VQE-PT with and without post-processing techniques, relative to the FCI. The shaded region indicates chemical accuracy. (b) Potential energy surfaces (PESs) computed using the FCI method, together with VQE and VQE-PT results with and without post-processing techniques.
}}
	\label{fig:result}
\end{figure}
\begin{table}[!htbp]
\caption{Mean absolute errors (MAE) of ground-state energies (in mHa) for the VQE and VQE-PT with symmetry verification (SV) and 2-RDM reconstruction, relative to FCI benchmark.}
\label{table4}
\centering
\small
\begin{tabular}{cccc}
       \toprule
       \midrule
         Method & MAE & Method & MAE\\
       \midrule
        VQE (Raw)& 89.8 & VQE-PT (Raw) & 23.8\\
        VQE +SV& 14.9 & VQE-PT +SV  & 7.6\\
        VQE +SV+RDM& 5.9 & VQE-PT +SV+RDM & 1.2\\
        CASCI (2e, 2o) &   {5.9}   & Noiseless VQE-PT &  1.2    \\
        \bottomrule  
    \end{tabular}
\end{table}

Figure~\ref{fig:result} and Table~\ref{table4} summarize the experimental results obtained from the Quafu quantum processor for both VQE and VQE-PT, evaluated under a hierarchy of post-processing strategies. In the absence of advanced post-processing (Raw: readout error mitigation only), standard VQE yielded a large mean absolute error (MAE) of 89.8 mHa due to device noise. VQE-PT inherently reduced this to 23.8 mHa via perturbative corrections, though noise remained significant. Applying SV drastically improved fidelity, reducing the MAE by factors of 6.0 (VQE) and 3.1 (VQE-PT), confirming the effectiveness of symmetry constraints. {{\color{blue}}The most significant gain was achieved after applying the $N$-representability-constrained 2-RDM reconstruction.} With this full SV+RDM strategy, the standard VQE yielded an MAE of 5.9 mHa, which matches the theoretical limit of the (2e, 2o) active-space CASCI (5.9 mHa). This indicates that VQE has successfully captured the static correlation but remained limited by the missing dynamic correlation from the external space. In stark contrast, the VQE-PT method achieved a remarkable MAE of 1.2 mHa, matching the noiseless algorithmic limit, falling well within chemical accuracy (see also Fig.~\ref{fig:result}(a)). The corresponding potential energy surface (PES) in Fig.~\ref{fig:result}(b) and its inset further demonstrate that VQE-PT accurately reproduces the dissociation curve, correcting the quantitative deviations observed in the standard VQE results. Additional experimental data are provided in the Supporting Information.

\begin{table}[!htbp]
\caption{Comparison of the average quantum resource requirements and the theoretical MAEs for the ground state of the F$_2$ molecule across PES. The MAEs are calculated relative to the FCI (18e, 10o) benchmark. All presented errors are obtained via exact noiseless numerical simulations to evaluate the algorithmic limits of each ansatz.}
\label{table2}
\centering
\small 
\setlength{\tabcolsep}{4pt}
\begin{tabular}{lccccc}
\toprule
\midrule
Method & VQE Space  & Qubits & Params  & CNOTs & MAE (mHa) \\
\midrule
UCCSD & (18e, 10o) & 20 & 21 & 453.7 & 0.00  \\
UCCSD & (14e, 8o) & 16& 11 & 205.4 & 0.02  \\
VQE-PT & (2e, 2o) & 4& 1 & 13 (UCCSD)/4 (This work) & 1.23 \\
\bottomrule 
\end{tabular}
\end{table}

Table \ref{table2} compares the theoretical quantum resource requirements and the mean energetic accuracy of our VQE-PT protocol against standard unitary coupled-cluster (UCCSD) baselines. To ensure a fair comparison, all MAEs in this table are evaluated via noiseless numerical simulations. The full-space UCCSD ansatz  requires an average of 453.7 CNOTs~\cite{yordanov2021qubit, yordanov2020efficient} to reach FCI-level energies, while a frozen-core UCCSD implementation reduces this to 205.4 CNOTs with a residual error of 0.02 mHa. Although theoretically accurate, these deep ansatze are practically unfeasible on current NISQ devices. Given typical two-qubit gate error rates on the order of $10^{-2}$, executing hundreds of CNOTs leads to an exponential decay of circuit fidelity that overwhelms the physical signal. Furthermore, deep circuits subject to realistic local noise can exhibit noise-induced barren plateaus~\cite{wang2021noise}, i.e., an exponential suppression of parameter gradients with system size or circuit depth — rendering the classical optimization step highly intractable.

For the VQE-PT calculation, applying a standard UCCSD ansatz to the (2e, 2o) active space requires only 13 CNOTs. The tailored ansatz used in our hardware experiment reduces the entanglement cost to 4 CNOTs. Because both ansatze prepare the exact same active-space reference state, they yield an identical noiseless VQE-PT MAE of 1.23 mHa, falling well within the chemical accuracy threshold ($\approx 1.6$ mHa). Crucially, this massive algorithmic resource reduction is what ultimately permits our robust hardware execution. This favorable resource-to-accuracy ratio demonstrates that VQE-PT is a highly practical framework, capable of effectively recovering missing dynamic correlation while operating strictly within the stringent coherence budgets of near-term quantum processors.

This dramatic reduction in the circuit complexity of the VQE-PT method offers two key algorithmic advantages. First, by restricting the variational optimization to a compact active space, the number of parameters is significantly reduced—for instance, from 11 down to just 1 in this specific case. This reduction simplifies the optimization landscape and effectively mitigates the severe convergence difficulties prevalent in the high-dimensional parameter spaces of large-scale VQE. Second, the approach exhibits enhanced noise resilience. The intrinsically shallow depth of the active-space circuit ensures that the reference wave function is prepared well within the coherence time of current NISQ devices. This minimizes the accumulation of gate errors prior to the application of the perturbative correction, enabling a high-fidelity baseline for recovering dynamic correlation.

Consequently, by strictly confining both the VQE procedure and the subsequent measurements to a compact active space, VQE-PT drastically reduces the required qubit count and circuit depth, empowering current quantum computers to tackle significantly larger molecular systems beyond the computational reach of standard full-space VQE.

\section{Conclusion} 
In conclusion, we have developed an RDM-based perturbative VQE framework for recovering dynamic correlation beyond a compact active space without increasing the qubit count or variational circuit depth. The framework comprises a fully coupled formulation, VQE-PTs, and a diagonal approximation, VQE-PT. Numerical simulations of HF, N$_2$, and F$_2$ show that VQE-PTs provides the more robust accuracy-oriented treatment, while VQE-PT offers an efficient approximation when the couplings among different internally contracted perturbers are weak. Both variants use the same RDM measurement data in the present implementation; VQE-PT simplifies the subsequent classical processing by avoiding the coupled amplitude solve and potential numerical instabilities during matrix inversion. We experimentally implemented this diagonal variant on the Quafu superconducting quantum processor for F$_2$, obtaining a mean absolute error of 1.2 mHa and achieving chemical accuracy across most of the potential energy surface. Together, these results establish perturbative VQE as a flexible framework for near-term calculations.

Although the perturbative correction effectively maintains a shallow circuit depth, it formally depends on evaluating high-order RDMs (up to 4-RDMs). As the active space grows, the sampling burden could become significant. To mitigate this, the sparsity of the Hamiltonian and the localized structure of typical excitations permit an \textit{a priori} screening strategy to identify the small subset of RDM elements actually required. Furthermore, integrating the perturbative VQE framework with advanced measurement techniques, such as classical shadows~\cite{huang2020predicting}, Pauli grouping~\cite{tilly2022variational,verteletskyi2020measurement,gokhale2019minimizing}, and cumulant approximations~\cite{harris2002cumulant, takemori2023balancing}, can substantially alleviate this overhead (see Supporting Information for details). It should be noted, however, that cumulant approximations may introduce truncation errors that require careful benchmarking.

An important direction for future work is the integration of the present perturbative VQE framework with orbital optimization. In classical electronic structure theory, orbital-optimized multireference methods, such as CASSCF and orbital-optimized perturbation theories, play a crucial role in achieving balanced descriptions of static and dynamic correlation~\cite{Roos1980,Olsen2011}. Recent studies have demonstrated that incorporating orbital optimization within variational quantum algorithms can significantly improve accuracy and convergence properties~\cite{mizukami2020orbital,10.1063/1.5141835,Yao2021,omiya2022analytical}. Extending perturbative VQE to an orbital-optimized framework would enable a self-consistent treatment in which both the wave function parameters and the underlying orbital basis are variationally refined, enabling a highly compact active space by absorbing orbital relaxation effects, thereby improving the efficiency and accuracy of subsequent perturbative corrections. 

\section*{Associated Content}
The Supporting Information includes additional experimental data and hardware parameters for the quantum processor; detailed descriptions of perturbative operators; proof of the maximum RDM rank required in the perturbative VQE framework; and analysis of the effect of basis-set expansion.

\section*{Acknowledgments}
This work was supported by the Innovation Program for Quantum Science and Technology (2021ZD0303306, 2023ZD0300200), the National Natural Science Foundation of China (22422304, 22393913, 22303090, 22303005, 12404560), the robotic AI-Scientist platform of the Chinese Academy of Sciences, and the Supercomputing Center of the University of Science and Technology of China. We would like to express our sincere gratitude to Prof. Yang Guo and Rahul Maitra for insightful discussions and valuable feedback on the methodology. We thank Dedong Wan for helpful discussions on classical benchmark calculations and related literature.

\section*{Conflicts of Interest}
There are no conflicts of interest to declare.

\clearpage
\newpage

\begin{center}
    \LARGE \textbf{Supporting Information for:} \\
    \vspace{0.5cm}
    \Large \textbf{Perturbative Variational Quantum Eigensolver via Reduced Density Matrices} \\
    \vspace{0.5cm}
    \normalsize Yuhan Zheng, Yibin Guo, Huili Zhang, Jie Liu, Xiongzhi Zeng, Xiaoxia Cai, Zhenyu Li, and Jinlong Yang
\end{center}

\setcounter{section}{0}
\setcounter{equation}{0}
\setcounter{figure}{0}
\setcounter{table}{0}
\setcounter{page}{1}

\renewcommand{\thesection}{S\arabic{section}}
\renewcommand{\theequation}{S\arabic{equation}}
\renewcommand{\thefigure}{S\arabic{figure}}
\renewcommand{\thetable}{S\arabic{table}}

\section{Experimental Details}

\subsection{Device Information}
The experiment was conducted on the Baihua chip, part of the Quafu quantum cloud platform. Baihua is a superconducting quantum processor with 156 quantum qubits and 172 tunable couplers. Four qubits, including Q57, Q58, Q59, and Q60, were utilized to perform quantum simulations. The basic parameters and gate fidelity of these four qubits are shown in Table~\ref{table1}.

\begin{table}[H]
\centering 
\caption{\centering Basic parameters and gate fidelity of qubits}
\label{table1}
\begin{tabular}{m{5cm} m{2cm} m{2cm} m{2cm} m{2cm} m{2cm} m{2cm} m{2cm} m{2cm}|}
\toprule
& \multicolumn{2}{c}{\textbf{Q57}} & \multicolumn{2}{c}{\textbf{Q58}} & \multicolumn{2}{c}{\textbf{Q59}} & \multicolumn{2}{c}{\textbf{Q60}} \\
\midrule
Qubit frequency(GHz) & \multicolumn{2}{c}{4.066} & \multicolumn{2}{c}{4.407} & \multicolumn{2}{c}{4.134} & \multicolumn{2}{c}{4.396} \\
Anharmonicity(GHz) & \multicolumn{2}{c}{0.196} & \multicolumn{2}{c}{ 0.178} & \multicolumn{2}{c}{0.193} & \multicolumn{2}{c}{0.181} \\
Readout frequency(GHz) & \multicolumn{2}{c}{6.754} & \multicolumn{2}{c}{7.063} & \multicolumn{2}{c}{6.799} & \multicolumn{2}{c}{7.109} \\
T1($\mu$s) & \multicolumn{2}{c}{63.9} & \multicolumn{2}{c}{63.9} & \multicolumn{2}{c}{82.7} & \multicolumn{2}{c}{62.4} \\
T2 Ramsey($\mu$s) & \multicolumn{2}{c}{35.8} & \multicolumn{2}{c}{21.4} & \multicolumn{2}{c}{19.6} & \multicolumn{2}{c}{28.3} \\
single-qubit gate fidelity(\%) & \multicolumn{2}{c}{99.93} & \multicolumn{2}{c}{99.93} & \multicolumn{2}{c}{99.94} & \multicolumn{2}{c}{99.93} \\
\multicolumn{2}{m{5.8cm}}{CZ gate fidelity(\%)} & \multicolumn{2}{c}{98.5} & \multicolumn{2}{c}{99.2} & \multicolumn{2}{c}{98.7} \\

\bottomrule
\end{tabular}
\end{table}

The compilation of the quantum circuit in this experiment adopted U3 gate decomposition as the standard method for single-qubit gate synthesis. During compilation, adjacent single-qubit gates were merged into a single equivalent unitary operation to reduce circuit depth. The native two-qubit gate on the Baihua chip is the controlled-Z (CZ) gate, which serves as the fundamental entangling operation on this superconducting hardware.

\subsection{Experimental Process}
\begin{figure}[H]
    \centering
    \includegraphics[width=1\linewidth]{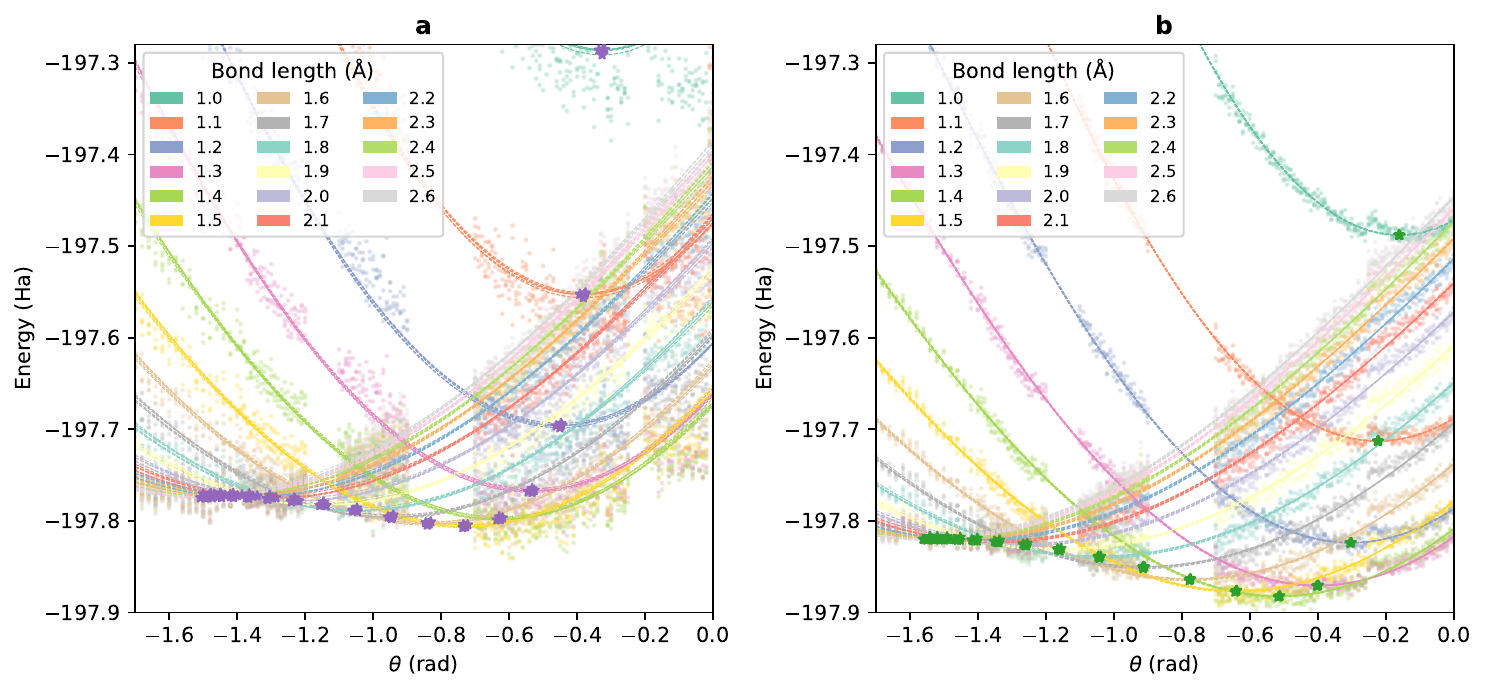}
    \caption{Potential-energy data for F$_2$ obtained from single-parameter sweeps of a variational quantum eigensolver (VQE) ansatz on quantum hardware. The plots show the measured energy expectation values (in hartree) as a function of the variational parameter $\theta$ for various F-F bond lengths. For each bond length, results from five independent experimental runs are shown. The scattered dots represent measurement outcomes, while the lines correspond to fitted curves used to locate the energy minima (marked by stars). (a) Raw experimental results corrected only for readout errors, where variations among the five runs are visible due to hardware noise. (b) Experimental results with additional symmetry verification post-processing. The fitted curves from the five independent runs show strong agreement.}
    \label{S1}
\end{figure}
Fig.~\ref{S1} presents the experimental energy profiles obtained during the variational quantum eigensolver (VQE) optimization for the F$_2$ molecule. Given the single-parameter structure of the ansatz (see Fig.~3(d) in the main text), the VQE optimization was implemented via a parameter sweep. To mitigate the impact of temporal fluctuations in device noise (parameter drift), we adopted a simultaneous measurement strategy. Instead of optimizing the VQE parameters first and subsequently measuring the perturbative terms, we evaluated the complete set of 22 Pauli strings required for both the VQE energy and the perturbative correction at every parameter point during the sweep.  This ensures that the reference state used for perturbation theory is fully consistent with the optimized VQE state. To minimize the hardware sampling overhead, observables were grouped into 9 distinct measurement bases. Specifically, all 14 purely $Z$-basis strings ($Z_0$, $Z_1$, $Z_2$, $Z_3$, $Z_0Z_1$, $Z_0Z_2$, $Z_0Z_3$, $Z_1Z_2$, $Z_1Z_3$, $Z_2Z_3$, $Z_0Z_1Z_2$, $Z_0Z_2Z_3$, $Z_1Z_2Z_3$, $Z_0Z_1Z_3$) were evaluated simultaneously from a single measurement setting in the computational basis ($Z_0Z_1Z_2Z_3$), while the remaining 8 terms ($Y_0X_1Y_2X_3$, $Y_0Y_1X_2X_3$, $X_0Y_1Y_2X_3$, $X_0X_1X_2X_3$, $Y_0Y_1Y_2Y_3$, $Y_0X_1X_2Y_3$, $X_0X_1Y_2Y_3$, $X_0Y_1X_2Y_3$) were measured individually. Independent experiments were repeated five times for each F-F bond length with 1024 shots per Pauli string.

The VQE energy landscape was reconstructed using the relevant subset of 14 Pauli strings ($Z_0$, $Z_1$, $Z_2$, $Z_3$, $Z_0Z_1$, $Z_0Z_2$, $Z_0Z_3$, $Z_1Z_2$, $Z_1Z_3$, $Z_2Z_3$,   $Y_0Y_1X_2X_3$, $X_0Y_1Y_2X_3$,   $Y_0X_1X_2Y_3$, $X_0X_1Y_2Y_3$). As shown in Fig.~\ref{S1}, the energy data from each run were fitted to identify the minimum-energy parameter (marked by stars). Because the variational parameter was sampled at discrete values during the sweep, the fitted minimum did not necessarily coincide with an explicitly sampled point. To mitigate discretization errors and statistical fluctuations, we averaged the energies from four parameter points closest to the fitted minimum to determine the final VQE energy; the perturbative corrections were then computed from these optimal points. The corresponding optimized parameters are summarized in Table~\ref{table4}. Fig.~\ref{S1}(a) displays the raw experimental results (corrected for readout errors~\cite{van2022model} only), where hardware noise causes fluctuations among the five runs. In contrast, after applying symmetry verification~\cite{PhysRevA.98.062339, sagastizabal2019experimental} (Fig.\ref{S1}(b)), the energy profiles from the five runs show strong agreement, indicating effective suppression of the variance induced by device noise, and yielding energies closer to the noiseless simulation.

Table~\ref{tab:exp_energies} summarizes the mean energies and standard deviations obtained from the five independent runs. Symmetry verification reduces the standard deviation, and RDM purification further suppresses residual statistical uncertainty. For all bond lengths and post-processing levels, the perturbative variational quantum eigensolver (VQE-PT) energies are consistently lower than standard VQE results, confirming that the perturbative correction recovers correlation effects beyond the active space, improving accuracy without increasing the depth or width of the quantum circuit.


\begin{table}[htbp]
\centering
\caption{Optimized parameters yielding the minimum energy at various bond lengths in the variational quantum eigensolver (VQE) calculations: raw experimental data (Raw VQE), symmetry-verified experimental data (VQE +SV), and noiseless classical simulation (Noiseless VQE). Experimental values represent the mean obtained from five independent runs.}
\label{table4}
\resizebox{\linewidth}{!}{
    \begin{tabular}{c|cccccccccc}
\toprule
\textbf{Bond (\AA)} & 1.0 & 1.1 & 1.2 & 1.3 & 1.4 & 1.5 & 1.6 & 1.7 & 1.8  \\
\midrule
\textbf{Raw VQE}  &-0.326085 & -0.382259 & -0.452657 & -0.536228 & -0.631276  & -0.735294 & -0.844635 & -0.954549 & -1.059850  \\
\textbf{VQE +SV}   &-0.160397 & -0.222984 & -0.303044 & -0.400295 & -0.513577 & -0.640279 & -0.775592 & -0.912497 & -1.043067\\
\textbf{Noiseless VQE} &  -0.137354 & -0.202056 & -0.284538 & -0.384217 & -0.499510 & -0.627307 & -0.762391 & -0.897629 & -1.025399\\
\bottomrule
\end{tabular}}

\vspace{0.5em}

\resizebox{\linewidth}{!}{
    \begin{tabular}{c|ccccccccc}
\toprule
\textbf{Bond (\AA)}  & 1.9 & 2.0 & 2.1 & 2.2 & 2.3 & 2.4 & 2.5 & 2.6 \\
\midrule
\textbf{Raw VQE}   & -1.155991 & -1.239955 & -1.310536 & -1.368065 & -1.413854 & -1.449664 & -1.477318 & -1.498498 \\
\textbf{VQE +SV}  & -1.160574 & -1.261030 & -1.343404  & -1.408835  & -1.459609  & -1.498343 & -1.527528 & -1.549312 \\
\textbf{Noiseless VQE}  & -1.139530 & -1.236623 & -1.315990 & -1.378936 & -1.427745 & -1.464976 & -1.493030 & -1.513974\\ 
\bottomrule
\end{tabular}}
\end{table}

{\scriptsize
\setlength{\tabcolsep}{4pt}
\begin{longtable}{
    p{0.7cm} |
    p{2.3cm} p{2.3cm} p{2.3cm} |
    p{2.3cm} p{2.3cm} p{2.3cm}
}
\caption{Experimental ground-state energies (in hartree) for the F$_2$ molecule. Mean energy and standard deviation ($\bar{E} \pm \sigma$) obtained from five independent experiments.}
\label{tab:exp_energies} \\

\toprule
\multicolumn{1}{c|}{} &
\multicolumn{3}{c|}{\textbf{Standard VQE}} &
\multicolumn{3}{c}{\textbf{VQE-PT}} \\
\cmidrule(lr){2-4} \cmidrule(lr){5-7}
 \textbf{R(\AA)} & \textbf{Raw} & \textbf{SV} & \textbf{SV+RDM}
 & \textbf{Raw} & \textbf{SV} & \textbf{SV+RDM} \\
\midrule
\endfirsthead

\toprule
\multicolumn{1}{c|}{} &
\multicolumn{3}{c|}{\textbf{Standard VQE}} &
\multicolumn{3}{c}{\textbf{VQE-PT}} \\
\cmidrule(lr){2-4} \cmidrule(lr){5-7}
 \textbf{R(\AA)}& \textbf{Raw} & \textbf{SV} & \textbf{SV+RDM}
 & \textbf{Raw} & \textbf{SV} & \textbf{SV+RDM} \\
\midrule
\endhead

\bottomrule
\endfoot

1.0 & $-197.289\pm0.009$ & $-197.486 \pm 0.002$ & $-197.501 \pm 0.000$ & $-197.416 \pm 0.012$ & $-197.499 \pm 0.002$ & $-197.514 \pm 0.000$ \\
1.1 & $-197.566 \pm 0.006$ & $-197.703 \pm 0.001$ & $-197.724 \pm 0.000$ & $-197.671 \pm 0.004$ & $-197.718 \pm 0.001$ & $-197.738 \pm 0.000$ \\
1.2 & $-197.713 \pm 0.003$ & $-197.826 \pm 0.002$ & $-197.833 \pm 0.000$ & $-197.806 \pm 0.005$ & $-197.841 \pm 0.002$ & $-197.847 \pm 0.000$ \\
1.3 & $-197.778 \pm 0.008$ & $-197.868 \pm 0.001$ & $-197.879 \pm 0.000$ & $-197.863 \pm 0.004$ & $-197.882 \pm 0.001$ & $-197.893 \pm 0.000$ \\
1.4 & $-197.804 \pm 0.007$ & $-197.884 \pm 0.002$ & $-197.890 \pm 0.000$ & $-197.880 \pm 0.004$ & $-197.897 \pm 0.002$ & $-197.903 \pm 0.000$ \\
1.5 & $-197.794 \pm 0.007$ & $-197.879 \pm 0.002$ & $-197.884 \pm 0.000$ & $-197.870 \pm 0.002$ & $-197.891 \pm 0.002$ & $-197.896 \pm 0.000$ \\
1.6 & $-197.794 \pm 0.007$ & $-197.862 \pm 0.002$ & $-197.872 \pm 0.000$ & $-197.860 \pm 0.001$ & $-197.871 \pm 0.002$ & $-197.882 \pm 0.000$ \\
1.7 & $-197.780 \pm 0.006$ & $-197.844 \pm 0.002$ & $-197.858 \pm 0.000$ & $-197.845 \pm 0.003$ & $-197.853 \pm 0.002$ & $-197.867 \pm 0.000$ \\
1.8 & $-197.776 \pm 0.004$ & $-197.838 \pm 0.003$ & $-197.847 \pm 0.000$ & $-197.839 \pm 0.003$ & $-197.844 \pm 0.003$ & $-197.854 \pm 0.000$ \\
1.9 & $-197.777 \pm 0.002$ & $-197.830 \pm 0.003$ & $-197.839 \pm 0.000$ & $-197.833 \pm 0.003$ & $-197.835 \pm 0.003$ & $-197.844 \pm 0.000$ \\
2.0 & $-197.779 \pm 0.007$ & $-197.834 \pm 0.005$ & $-197.833 \pm 0.000$ & $-197.829 \pm 0.004$ & $-197.837 \pm 0.005$ & $-197.837 \pm 0.000$ \\
2.1 & $-197.775 \pm 0.004$ & $-197.821 \pm 0.002$ & $-197.830 \pm 0.000$ & $-197.824 \pm 0.004$ & $-197.823 \pm 0.002$ & $-197.833 \pm 0.000$ \\
2.2 & $-197.772 \pm 0.003$ & $-197.819 \pm 0.005$ & $-197.828 \pm 0.000$ & $-197.819 \pm 0.003$ & $-197.821 \pm 0.005$ & $-197.830 \pm 0.000$ \\
2.3 & $-197.772 \pm 0.005$ & $-197.819 \pm 0.004$ & $-197.827 \pm 0.000$ & $-197.817 \pm 0.005$ & $-197.821 \pm 0.004$ & $-197.828 \pm 0.000$ \\
2.4 & $-197.774 \pm 0.006$ & $-197.820 \pm 0.003$ & $-197.826 \pm 0.000$ & $-197.815 \pm 0.005$ & $-197.821 \pm 0.003$ & $-197.827 \pm 0.000$ \\
2.5 & $-197.778 \pm 0.003$ & $-197.820 \pm 0.003$ & $-197.826 \pm 0.000$ & $-197.816 \pm 0.005$ & $-197.821 \pm 0.003$ & $-197.827 \pm 0.000$ \\
2.6 & $-197.779 \pm 0.004$ & $-197.820 \pm 0.003$ & $-197.826 \pm 0.000$ & $-197.817 \pm 0.004$ & $-197.820 \pm 0.003$ & $-197.827 \pm 0.000$ \\
\end{longtable}
}

\section{Evaluation of the Perturbative VQE Framework}\label{SI:VQE_PT_Details}

\subsection{Perturbative VQE theory}
As introduced in the main text, the reference wave function of a ground state for the VQE-PT method is defined as the tensor product of the inactive, active, and virtual spatial components:
\begin{equation}\label{eq:psi_0}
    | \Psi_0 \rangle  = \rm|\Psi_{inact}\rangle\otimes|\Psi_{act}(\boldsymbol{\theta}_0)\rangle\otimes|\Psi_{vir}\rangle,
\end{equation}
where $\rm|\Psi_{inact}\rangle$ and $\rm|\Psi_{vir}\rangle$ represent doubly occupied and unoccupied (vacuum) states in the inactive and virtual spaces, respectively. The active-space wave function $\rm|\Psi_{act}\rangle$ is prepared using the adaptive derivative-assembled pseudo-Trotter (ADAPT) VQE~\cite{grimsley2019adaptive} algorithm. The ADAPT-VQE calculations employed a spin-adapted operator pool (see Section \ref{sec:adapt}) consisting of occupied-to-virtual single and double excitation operators within the active space. For the non-degenerate ground states considered in this work, the Hartree-Fock reference belongs to the same irreducible representation as the target state. Therefore, only totally symmetric excitation operators were retained to preserve the spatial symmetry of the wavefunction during the ADAPT-VQE optimization. At each ADAPT iteration, the operator with the largest gradient magnitude was selected and appended to the ansatz. The iterative procedure was terminated when the norm of the operator-gradient vector fell below $10^{-4}$. Furthermore, at each macroscopic ADAPT step, the VQE parameter optimization was performed using the BFGS algorithm, terminating when the gradient norm fell below $10^{-5}$. 

To recover the dynamic correlation energy missing from the active-space simulation, the first-order wave function $| \Psi^{(1)} \rangle$ is expanded within a linear space spanned by a set of $K$ internally-contracted~\cite{werner1988efficient} perturbers, denoted as $\{|\Phi_I\rangle\}_{I=1}^K$. These perturbers are generated by applying spin-adapted excitation operators $\hat{r}_I$~\cite{helgaker2013molecular, chan2021molecular} (see Section \ref{sec:adapt}) directly to the reference state, i.e., $|\Phi_I\rangle=\hat{r}_I|\Psi_0\rangle$. 
$\{\ket{\Phi_I}\}$ is generally non-orthogonal and potentially linearly dependent ($S_{IJ} = \braket{\Phi_I|\Phi_J} \neq \delta_{IJ}$).

We construct an orthonormal basis $\{|\Psi_J \rangle\}$ for the perturbative subspace via symmetric (Löwdin) orthogonalization~\cite{mayer2002lowdin, lowdin1950non}:
\begin{equation}\label{eq:lowdin}
|\Psi_J \rangle = \sum_{I \neq 0}^{K} {\mathbf{S}^{-\frac{1}{2}}_{IJ}}\ket{\Phi_I},
\end{equation}
where the reference state $|\Psi_0 \rangle$ is excluded from the summation as it is already orthogonal to all perturbers $\ket{\Phi_I}$. The dimension of this orthogonalized subspace is denoted as $K'$ ($K' \le K$).

With the well-defined orthonormal basis $\{|\Psi_0\rangle, |\Psi_1\rangle, \dots, |\Psi_{K'}\rangle\}$, we apply the Rayleigh-Schrödinger (RS) PT~\cite{sakurai2020modern} formalism by partitioning the exact Hamiltonian as $\hat{H} = \hat{H}_0 + \hat{V}$. Specifically, we adopt the diagonal partitioning scheme, where the zeroth-order Hamiltonian $\hat{H}_0$ is defined as the diagonal part of the full Hamiltonian in this basis:
\begin{equation}\label{eq:H0}
    \hat{H}_0 = E_{\rm VQE} |\Psi_0\rangle\langle\Psi_0| + \sum_{I=1}^{K'} \tilde{H}_{II} |\Psi_I\rangle\langle\Psi_I|,
\end{equation}
where $E_{\rm VQE} = \langle \Psi_0|\hat{H}|\Psi_0\rangle$ and $\tilde{H}_{II} = \langle \Psi_I|\hat{H}|\Psi_I\rangle$ are the exact expectation values of the respective states. The perturbation operator is simply $\hat{V} = \hat{H} - \hat{H}_0$. 

By construction, the reference state and the orthogonalized perturbers are exact eigenstates of $\hat{H}_0$ with zeroth-order energies $E^{(0)} = E_{\rm VQE}$ and $E_I^{(0)} = \tilde{H}_{II}$, respectively. Under the standard Rayleigh-Schrödinger perturbation theory (RSPT)~\cite{sakurai2020modern} framework, the first-order energy correction vanishes ($E^{(1)} = \langle\Psi_0|\hat{V}|\Psi_0\rangle = 0$), and the dynamic correlation energy is directly obtained from the second-order correction:
\begin{equation}\label{eq:PT2}
    \Delta E_{\rm VQE-PT} = \sum_{I=1}^{K'} \frac{|\langle \Psi_I | \hat{V} | \Psi_0 \rangle|^2}{E^{(0)} - E_I^{(0)}} = \sum_{I=1}^{K'} \frac{|\langle \Psi_I | \hat{H} | \Psi_0 \rangle|^2}{E_{\rm VQE} - \tilde{H}_{II}}.
\end{equation}

The fully coupled VQE-PTs formulation uses the same reference state, perturber space, overlap matrix, and Hamiltonian matrix elements, but retains the complete Hamiltonian in the orthonormal perturber basis.

Its amplitudes are obtained from
\begin{equation}\label{eq:PTs_linear_system}
\sum_{J=1}^{K'}\left(\tilde{H}_{IJ}-E_{\rm VQE}\delta_{IJ}\right)\tilde{c}_J
=-\langle\Psi_I|\hat{H}|\Psi_0\rangle,
\end{equation}
and the energy correction is $\Delta E_{\rm VQE-PTs}=\sum_I\tilde{c}_I\langle\Psi_0|\hat{H}|\Psi_I\rangle$. Thus, VQE-PT and VQE-PTs differ in the classical treatment of $\hat{H}_0$, not in the quantum-measured matrix elements required to construct it.

The perturbative correction relies on the evaluation of transition matrix elements $\langle \Psi_{I}|\hat{H}|\Psi_{0}\rangle$. For an exact active-space eigenstate, the Hamiltonian matrix elements between the reference and configurations confined entirely within the active space vanish. For the converged ADAPT-VQE states used here, these matrix elements are numerically negligible within the optimization tolerance. 
\begin{equation}
\langle \Psi_0 | \hat{r}_I^\dagger \hat{H} | \Psi_0 \rangle
\approx 0,
\end{equation}
where $\hat r_I$ acts only within the active space. 

\subsection{Purely Active-Space Excitations and the Perturbative Model Space}
\label{sec:active_space}

The second-order perturbative energy correction requires evaluating the transition matrix elements
\begin{equation}
\langle \Psi_I|\hat H|\Psi_0\rangle,
\end{equation}
where the perturbers are generated via $|\Psi_I\rangle=\hat r_I|\Psi_0\rangle$. 
To formulate a rigorous perturbation theory, the perturbers must be strictly orthogonal to the reference state, i.e., $\langle \Psi_0 | \hat{r}_I | \Psi_0 \rangle = 0$. For excitations involving inactive or virtual orbitals, this orthogonality is naturally satisfied due to the exact occupations of the external spaces. However, for excitations confined entirely within the active space, standard excitation operators do not guarantee orthogonality. Therefore, if purely active-space perturbers were to be considered, they must be constructed using spin-adapted \textit{anti-Hermitian} excitation operators (which possess strictly vanishing expectation values for real wavefunctions). 

In the present work, purely active-space excitations are excluded from the perturbative operator pool for both VQE-PT and VQE-PTs. This follows from the active/external partition underlying the framework. Let $\hat P_{\rm act}$ project onto the space with the inactive and virtual occupations fixed at their reference values and arbitrary configurations within the active orbitals, and let $\hat Q_{\rm ext}=1-\hat P_{\rm act}$. Correlation within $\hat P_{\rm act}$ is assigned to the VQE active-space problem, whereas the perturbative correction is constructed exclusively from the externally excited sector $\hat Q_{\rm ext}$. Purely active-space excitations therefore belong to the VQE model space rather than to the perturbative space. The following analysis verifies that their direct couplings to the converged reference are negligible and quantifies their individual VQE-PT corrections.

Let $|\Psi_{\rm act}\rangle$ be the fully optimized active-space wave function. We can evaluate the energy gradient for a parameterized test state $|\Phi(\theta)\rangle = e^{\theta\hat{r}_u}|\Psi_{\rm act}\rangle$ generated by an active-space excitation operator $\hat{r}_u$. The gradient evaluated at the converged state is:
\begin{equation}
g_u = \left. \frac{\partial}{\partial\theta} \langle \Psi_{\rm act} | e^{\theta\hat r_u^\dagger} \hat H_{\rm act} e^{\theta\hat r_u} | \Psi_{\rm act} \rangle \right|_{\theta=0} = \langle \Psi_{\rm act} | \hat r_u^\dagger \hat H_{\rm act} + \hat H_{\rm act} \hat r_u | \Psi_{\rm act} \rangle.
\end{equation}
By construction, the excitation operators in ADAPT-VQE are anti-Hermitian ($\hat r_u^\dagger=-\hat r_u$). Utilizing this property along with the Hermiticity of $\hat H_{\rm act}$, the gradient expression simplifies to:
\begin{equation}
g_u = \langle \Psi_{\rm act}|[\hat H_{\rm act},\hat r_u]|\Psi_{\rm act}\rangle = 2\,{\rm Re}\left(\langle\Psi_{\rm act}|\hat H_{\rm act}\hat r_u|\Psi_{\rm act}\rangle\right).
\end{equation}

The ADAPT-VQE optimization explicitly terminates when the norm of the gradient vector satisfies $\left( \sum_u |g_u|^2 \right)^{1/2} < \epsilon$, where $\epsilon = 10^{-4}$ in the present work. This convergence criterion strictly enforces that
\begin{equation}
\left|{\rm Re}\left(\langle\Psi_{\rm act}|\hat H_{\rm act}\hat r_u|\Psi_{\rm act}\rangle\right)\right| < \epsilon/2
\end{equation}
for every internal active-space excitation operator. 

Using the tensor-product structure of the full reference state, $|\Psi_0\rangle = |\Psi_{\rm inact}\rangle \otimes |\Psi_{\rm act}\rangle \otimes |\Psi_{\rm vir}\rangle$, and noting that $\hat r_u$ acts exclusively inside the active space, the full-space transition matrix element reduces exactly to the active-space expectation value:
\begin{equation}
\left|{\rm Re}\left( \langle \Psi_0|\hat H\hat r_u|\Psi_0\rangle \right)\right| = \left|{\rm Re}\left( \langle\Psi_{\rm act}|\hat H_{\rm act}\hat r_u|\Psi_{\rm act}\rangle \right)\right| < \epsilon/2.
\end{equation}
Therefore, purely active-space excitations possess numerically negligible Hamiltonian couplings to the reference state once the ADAPT-VQE optimization has converged. These couplings are the numerator terms in VQE-PT and the source-vector components on the right-hand side of the VQE-PTs linear system Eq.~\eqref{eq:PTs_linear_system}. For VQE-PTs, however, a negligible source component alone would not prove that appending an internal state to an enlarged coupled system has no indirect effect through off-diagonal couplings. Accordingly, the exclusion of such states from VQE-PTs follows from the active/external model-space definition above, rather than from an assumption that all indirect couplings vanish.

To verify this theoretical property numerically across the entire dissociation profile, we explicitly constructed all purely active-space excitations and evaluated their perturbative contributions for the molecular systems considered in this work. We conducted this validation for both the decoupled VQE-PT scheme and the fully coupled VQE-PTs scheme.

For the diagonal VQE-PT approximation, we computed the transition matrix element $M_u = \langle\Psi_0|\hat H\hat r_u|\Psi_0\rangle$, the perturbative energy denominator $\Delta_u = E_{\rm VQE} - \tilde{H}_{II}$, and the resulting individual second-order energy correction $E_u^{(2)}= |M_u|^2 / \Delta_u$ for each active-space operator $\hat{r}_u$ across all bond lengths.

\begin{table}[h]
\centering
\renewcommand{\arraystretch}{1.3}
\caption{Maximum Hamiltonian transition couplings ($\max |M_u|$), minimum perturbative energy denominators ($\min |\Delta_u|$), and maximum individual diagonal VQE-PT contributions ($\max |E_u^{(2)}|$) associated with purely active-space excitation operators across the entire potential energy surfaces. All energy values are in hartree.}
\begin{tabular}{lccc}
\toprule
System & $\max |M_u|$ & $\min |\Delta_u|$ & $\max |E_u^{(2)}|$ \\
\midrule
HF     & $4.52\times10^{-6}$ & $0.628$ & $2.78\times10^{-11}$ \\
F$_2$  & $4.75\times10^{-6}$ & $0.752$ & $2.93\times10^{-11}$ \\
N$_2$  & $4.07\times10^{-5}$ & $0.303$ & $1.12\times10^{-9}$ \\
\bottomrule
\end{tabular}
\label{tab:active_validation}
\end{table}

As summarized in Table~\ref{tab:active_validation}, although the largest Hamiltonian coupling in \ce{N2} approaches the order of our gradient convergence threshold, the corresponding energy denominators remain substantially separated from zero (with a minimum energy gap of $0.303$ hartree). Consequently, the maximum individual diagonal VQE-PT contribution is strictly bounded by $1.12\times10^{-9}$ hartree. For \ce{HF} and \ce{F2}, these contributions are on the order of $10^{-11}$ hartree. These values are at least six orders of magnitude smaller than the target chemical accuracy ($1.6\times10^{-3}$ hartree), validating their numerical insignificance in VQE-PT.

For the fully coupled VQE-PTs approach, a vanishing source vector $\tilde{V}_I$ alone does not mathematically guarantee the absence of indirect correlation effects via off-diagonal Hamiltonian couplings ($\tilde{H}_{IJ}$) between internal and external perturbers. To rigorously quantify this indirect effect, we re-solved the complete VQE-PTs linear system (Eq.~\eqref{eq:PTs_linear_system}) by artificially re-incorporating all purely active-space perturbers into the interacting space. By comparing the resulting potential energy surfaces to the standard VQE-PTs results, we found that the maximum deviation across all sampled geometries was entirely negligible: less than $ 2.5\times 10^{-13}$ hartree for \ce{HF}, $3\times 10^{-11}$ hartree for \ce{F2}, and $2.4\times 10^{-4}$ hartree for the \ce{N2} molecule. All deviations fall well below chemical accuracy.

\subsection{Construction of the Perturbative Operator Pool}
Based on the analysis above, only excitations that couple the active space to the external spaces are included in the perturbative operator pool. Specifically, we consider excitations from inactive to active/virtual orbitals, and from active to virtual orbitals. The operator pool comprises the following spin-adapted single and double excitations: 
\begin{equation}
\{\hat{r}_I\}=\{\hat{r}_i^u,\allowbreak\,\hat{r}_i^a,\allowbreak\,\hat{r}_u^a,\allowbreak\,\hat{r}_{ij}^{uv},\,\hat{r}_{ij}^{ab},\allowbreak\,\hat{r}_{uv}^{ab},\,\hat{r}_{ iu}^{vw},\allowbreak\,\hat{r}_{ iu}^{a v},\allowbreak\,\hat{r}_{uv}^{a w},\allowbreak\,\hat{r}_{ij}^{a u},\allowbreak\,\hat{r}_{u i}^{ab}\},
\end{equation}
where indices $i,j$ label inactive orbitals, $u,v,w$ label active orbitals, and $a,b$ label virtual orbitals. Indices $p,q,r,s$ will denote general orbitals spanning the entire orbital space. All these indices refer to spatial orbitals when constructing the spin-adapted excitation generators, but will denote individual spin orbitals during the evaluation of matrix elements and RDMs in the subsequent sections.

\subsection{Explicit Evaluation of Perturbative Corrections via Active-Space RDMs}\label{RDMs Requirement}
The perturbative correction can be calculated using RDMs in the active space. The $k$-RDM is defined as
\begin{equation}
^k D^{p_1p_2\dots p_k}_{q_1q_2\dots q_k} = \frac{1}{k!} \bra{\Psi_{\rm{act}}}\hat{a}_{p_1}^\dagger \hat{a}_{p_2}^\dagger\dots \hat{a}_{p_k}^\dagger \hat{a}_{q_k} \hat{a}_{q_{k-1}}\dots \hat{a}_{q_1}\ket{\Psi_{\rm{act}}}
\end{equation}

As discussed in the main text, both VQE-PT and VQE-PTs can be evaluated from the same active-space RDM data, requiring RDMs up to the fourth order. Here we provide the explicit proof. The overlap matrix $S_{IJ}=\langle\Psi_0|\hat r_I^\dagger\hat r_J|\Psi_0\rangle$ requires at most the 3-RDM because each retained perturber contains at least one external index. It therefore remains to bound the RDM ranks required for the reference-coupling vector and the perturber Hamiltonian matrix.

1. Reference-Coupling Vector

The perturbative numerator involves the coupling between the reference and the excited states, $\langle \Psi_{0}|\hat{r}_I^\dagger\hat{H}|\Psi_{0}\rangle$. Because the spin-adapted excitation operators $\hat{r}_I$ are strictly defined as linear combinations of elementary spin-orbital excitations $\hat{c}_I$ of the same excitation rank (as defined in Section \ref{sec:adapt}), the required maximum order of the active-space RDM is entirely determined by its individual constituent excitations. For simplicity of the algebraic derivation, we hereafter analyze the RDM requirements using a generic elementary component $\hat{c}_I$.

Since valid excitation operators $\hat{c}_I$ involve at least one index from the perturbative space, evaluating $\langle \Psi_{0}|\hat{c}_I^\dagger\hat{H}|\Psi_{0}\rangle$ requires contracting the two-body Hamiltonian with these specific indices of $\hat{c}_I$. This projection defines an effective operator acting solely on the active space:
\begin{equation}
\langle \Psi_{0}|\hat{c}_I^\dagger\hat{H}|\Psi_{0}\rangle = \bra{\Psi_{\rm act}}
\hat{O}_{\rm act}\ket{\Psi_{\rm act}},
\end{equation}
where $ \hat{O}_{\rm act} = \left( \bra{\Psi_\text{inact}}\otimes \bra{\Psi_\text{vir}} \right)   \hat{c}_I^\dagger\hat{H} \left( \ket{\Psi_\text{inact}} \otimes \ket{\Psi_\text{vir}} \right)$. 
After contracting the inactive or virtual indices (e.g., $i$ in $\hat{c}_I = \hat{a}_{u}^\dagger \hat{a}_{v}^\dagger \hat{a}_{w}\hat{a}_{i}$ or $a$ in $\hat{c}_I = \hat{a}_{a}^\dagger \hat{a}_{u}^\dagger \hat{a}_{v}\hat{a}_{w}$), the resulting effective operator $\hat{O}_{\rm act}$ involves at most three creation and three annihilation operators within the active space. Consequently, the evaluation of the perturbative numerator depends on active-space RDMs up to the third order (3-RDM). 
For example, evaluating the transition matrix element for $\hat{c}_I = \hat{a}_{u}^\dagger \hat{a}_{v}^\dagger \hat{a}_{w}\hat{a}_{i}$ with the two-body term $\frac{1}{2}\sum_{pqrs} g_{pqrs}(\hat{a}_{p}^\dagger \hat{a}_{q}^\dagger \hat{a}_{r} \hat{a}_{s})$ in the $\hat{H}$ yields: 
\begin{equation}
\begin{split}
  V^{u v}_{w i} &= \bra{ \rm \Psi_{0}}({\hat{a}_{u}^\dagger \hat{a}_{v}^\dagger \hat{a}_{w}\hat{a}_{i}})^\dagger \frac{1}{2}\sum_{pqrs} g_{pqrs}(\hat{a}_{p}^\dagger \hat{a}_{q}^\dagger \hat{a}_{r} \hat{a}_{s}) \ket{\rm \Psi_{0}}\\
  &= \sum_{xyz} (g_{xyz i}-g_{xyi z})[\frac{1}{2}(\delta_{ux}\delta_{vy}-\delta_{vx}\delta_{uy}){}^1D^{w}_{z} - \delta_{ux}{}^2D^{wy}_{vz} \\ &\quad+ \delta_{vx}{}^2D^{wy}_{uz}+\delta_{uy}{}^2D^{wx}_{vz}-\delta_{vy}{}^2D^{wx}_{uz}+3{}^3D^{wxy}_{vuz}] 
\end{split}
\end{equation}
and similarly for the virtual-space excitation:
\begin{equation}
\begin{split}
 V_{vw}^{a u} &= \bra{ \rm \Psi_{0}}({\hat{a}_{a}^\dagger \hat{a}_{u}^\dagger \hat{a}_{v}\hat{a}_{w}})^\dagger \frac{1}{2}\sum_{pqrs} g_{pqrs}(\hat{a}_{p}^\dagger \hat{a}_{q}^\dagger \hat{a}_{r} \hat{a}_{s}) \ket{\rm \Psi_{0}}\\
&= \sum_{xyz} (g_{axyz}-g_{xayz})(\delta_{ux}{}^2D^{wv}_{yz} -  3{}^3D^{wvx}_{uyz})
\end{split}
\end{equation}

2. Perturber Hamiltonian Matrix

Both variants require the Hamiltonian matrix in the nonorthogonal perturber basis,
\begin{equation}
H_{IJ}=\langle\Phi_I|\hat H|\Phi_J\rangle
=\langle\Psi_0|\hat r_I^\dagger\hat H\hat r_J|\Psi_0\rangle.
\end{equation} VQE-PT uses its diagonal elements $\tilde H_{II}$, but obtaining these elements generally still requires the full $H_{IJ}$ within each L\"owdin block. VQE-PTs uses the complete $\tilde H_{IJ}$ in Eq.~\eqref{eq:PTs_linear_system}. Consequently, it is sufficient to establish the maximum RDM rank required for a general matrix element $H_{IJ}$; the same bound then applies to both variants.

Evaluating this expectation value naively suggests a prohibitively high computational cost. By linearity, evaluating $\langle \Psi_0 | \hat{r}_I^\dagger \hat{H} \hat{r}_J | \Psi_0 \rangle$ reduces to evaluating matrix elements between their elementary components, $\langle \Psi_0 | \hat{c}_I^\dagger \hat{H} \hat{c}_J | \Psi_0 \rangle$. The full Hamiltonian $\hat{H}$ contains bare purely active terms ($\hat{A}_{\text{act}}$), mixed terms (containing both active and non-active indices), and purely external terms. For the mixed terms, the contraction between the external indices of $\hat{H}$ and the excitation operators inherently reduces the number of active-space operators involved, requiring at most the 4-body active-space RDM (4-RDM).

The primary computational bottleneck lies in the two-body term in $\hat{A}_{\text{act}}$, i.e., 
\begin{equation}
\langle \Psi_0 | \hat{c}_I^\dagger (\frac{1}{2}\sum_{uvwx} g_{uvwx} \hat{a}^\dagger_u \hat{a}^\dagger_v \hat{a}_w \hat{a}_x) \hat{c}_J | \Psi_0 \rangle.    
\end{equation}
 Taking semi-internal excitations as an example, the excitation operators can be factorized into active and non-active components, e.g., $\hat{c}_I = \hat{\tau}_I \hat{a}_{i}$ and $\hat{c}_J = \hat{\tau}_J \hat{a}_{j}$, where $\hat{\tau}$ involves exactly 3 active operators. Because the two-body term in $\hat{A}_{\text{act}}$ ($\frac{1}{2}\sum_{uvwx} g_{uvwx} \hat{a}^\dagger_u \hat{a}^\dagger_v \hat{a}_w \hat{a}_x$) acts exclusively within the active space, the overall expectation value factorizes across the orbital subspaces. The non-active operators ($\hat{a}_i^\dagger$ from $\hat{c}_I^\dagger$ and $\hat{a}_j$ from $\hat{c}_J$) can be grouped together by passing through the active-space operators to act directly on the non-active reference state. Given the closed-shell nature of the inactive reference $|\Psi_{\text{inact}}\rangle$, their contraction yields a Kronecker delta, $\langle \Psi_{\text{inact}} | \hat{a}_i^\dagger \hat{a}_j | \Psi_{\text{inact}} \rangle = \delta_{ij}$. Consequently, this expectation value is non-vanishing only if the spectator external indices perfectly match ($i=j$). Upon integrating out these external indices, the evaluation of the purely active-space component reduces to:
\begin{equation}\label{eq:bare_2b}
\Omega_{\text{bare-2b}} = \langle \Psi_\text{act} | \hat{\tau}_I^\dagger \left( \frac{1}{2}\sum_{uvwx} g_{uvwx} \hat{a}^\dagger_u \hat{a}^\dagger_v \hat{a}_w \hat{a}_x \right) \hat{\tau}_J | \Psi_\text{act} \rangle.
\end{equation}
A direct operator counting reveals that $\hat{\tau}_I^\dagger$ (3 active-space operators), the two-body part of $\hat{A}_{\text{act}}$ (4 active-space operators), and $\hat{\tau}_J$ (3 active-space operators) together constitute a composite string of up to 10 active-space operators. A brute-force evaluation of this term would seemingly demand the highly expensive 5-RDM.

To circumvent this, we consider the effective active-space Hamiltonian, formally defined as:
 \begin{equation}
     \hat{H}_{\rm act} = \left( \bra{\Psi_\text{inact}}\otimes \bra{\Psi_\text{vir}} \right) \hat{H} \left( \ket{\Psi_\text{inact}} \otimes \ket{\Psi_\text{vir}} \right),
 \end{equation}
 where $\rm|\Psi_{inact}\rangle$ and $\rm|\Psi_{vir}\rangle$ represent doubly occupied and vacuum states, respectively. $\hat{H}_{\rm act}$ consists of a bare two-body term (which is exactly identical to the one in $\hat{A}_{\text{act}}$), a mean-field dressed one-body term ($\sum_{uv}h^\text{eff}_{uv} \hat{a}^\dagger_u \hat{a}_v$), and a scalar core-energy term ($E_\text{core}$). Consequently, the bare two-body operator can be equivalently expressed by rearranging the terms:
\begin{equation}
     \frac{1}{2}\sum_{uvwx} g_{uvwx} \hat{a}^\dagger_u \hat{a}^\dagger_v \hat{a}_w \hat{a}_x = \hat{H}_{\rm act} - \sum_{uv}h^\text{eff}_{uv} \hat{a}^\dagger_u \hat{a}_v - E_\text{core}.
 \end{equation}
 Substituting this back into Eq.~\eqref{eq:bare_2b}, the target expectation value is decomposed into three parts. Evaluating the terms involving the dressed one-body operator and the scalar core-energy requires only up to the 4-RDM (e.g., $3 + 2 + 3 = 8$ operators) and 3-RDM, respectively. Therefore, our focus on reducing the 5-RDM bottleneck strictly narrows down to evaluating the $\hat{H}_{\rm act}$ component:
 \begin{equation}
 \Omega_{\rm act} = \langle \Psi_\text{act} | \hat{\tau}_I^\dagger \hat{H}_\text{act} \hat{\tau}_J | \Psi_\text{act} \rangle.
 \end{equation}
Fortunately, the required RDM rank for this term can be systematically reduced by utilizing the commutator relation $\hat{A}\hat{B} =[\hat{A}, \hat{B}] + \hat{B}\hat{A}$. We rewrite the target expression as:
\begin{equation}\label{eq:commutator_expansion}
\Omega_{\rm act} = \langle \Psi_\text{act} | \hat{\tau}^\dagger_{I} [\hat{H}_\text{act}, \hat{\tau}_{J}] | \Psi_\text{act} \rangle + \langle \Psi_\text{act} | \hat{\tau}^\dagger_{I} \hat{\tau}_{J} \hat{H}_{\text{act}} | \Psi_\text{act} \rangle.
\end{equation}
For an exact active-space eigenstate, $\hat{H}_{\text{act}} | \Psi_\text{act} \rangle = E_{\text{VQE}} | \Psi_\text{act} \rangle$. This is the active-space eigenstate assumption used in the formal RDM-rank analysis and is satisfied to the numerical convergence tolerance by the ADAPT-VQE states employed here. Under this assumption, the second term in Eq.~\eqref{eq:commutator_expansion} simplifies to $E_{\rm VQE} \langle \Psi_\text{act} | \hat{\tau}^\dagger_{I} \hat{\tau}_{J} | \Psi_\text{act} \rangle$. Since this term only involves the product $\hat{\tau}^\dagger_{I} \hat{\tau}_{J}$, it contains at most 6 active-space operators and therefore requires only up to the 3-RDM. For a reference with a finite active-space eigenstate residual, the equality acquires a residual-dependent correction that is neglected consistently with the VQE convergence tolerance in the present formulation.

For the first term, the commutator $[\hat{H}_{\text{act}}, \hat{\tau}_{J}]$ plays a crucial role in dimension reduction. According to Wick's theorem, the product of two second-quantized operator strings can be expanded into a fully normal-ordered term (representing zero contractions) plus a sum of terms containing at least one contraction:
\begin{equation}
\hat{A}\hat{B} = N[\hat{A}\hat{B}] + \{\text{normal-ordered contracted terms}\},
\end{equation}
where $N[\dots]$ denotes the normal ordering. Applying this expansion to the commutator yields:
\begin{equation}\label{eq:wick_commutator}[\hat{H}_{\text{act}}, \hat{\tau}_{J}] = \Big( N[\hat{H}_{\text{act}}\hat{\tau}_{J}] - N[\hat{\tau}_{J}\hat{H}_{\text{act}}] \Big) + \{\text{normal-ordered contracted terms}\}.
\end{equation}
Since the Hamiltonian $\hat{H}_{\text{act}}$ consists of an even number of fermionic operators (it conserves particle number), commuting it entirely past $\hat{\tau}_{J}$ inside the normal ordering requires an even number of fundamental fermion permutations. This accumulated permutation yields a strictly positive sign, ensuring that $N[\hat{H}_{\text{act}}\hat{\tau}_{J}] = N[\hat{\tau}_{J}\hat{H}_{\text{act}}]$. Therefore, the uncontracted terms in Eq.~\eqref{eq:wick_commutator} exactly cancel each other out.

Consequently, the commutator $[\hat{H}_{\text{act}}, \hat{\tau}_{J}]$ is non-vanishing \textit{only if} there is at least one Wick contraction between the creation and annihilation operators of $\hat{H}_{\text{act}}$ and $\hat{\tau}_{J}$. Each such contraction, governed by the anti-commutation relation $\{a_x, a_y^\dagger\} = \delta_{xy}$, replaces a pair of fermionic operators with a scalar Kronecker delta, effectively removing two operators from the total count. As a result, the maximum rank of the composite active-space operator $\hat{\tau}^\dagger_{I}[\hat{H}_{\text{act}}, \hat{\tau}_{J}]$ is strictly reduced from the naive count of 10 down to at most 8 operators. We therefore conclude that every element of the full perturber Hamiltonian matrix requires at most the 4-RDM. Together with the 3-RDM bounds for the overlap matrix and reference-coupling vector, this proves that both VQE-PT and VQE-PTs require active-space RDMs of at most fourth order.

\subsection{Strategies for Scalable RDM Evaluation}

While direct Pauli string measurements based on qubit-wise commutativity are highly effective for the molecules evaluated in this work, scaling either variant of the perturbative VQE framework to larger and more complex active spaces requires mitigating the $\mathcal{O}(N_{\text{act}}^8)$ measurement overhead associated with the 4-RDM, where $N_{\text{act}}$ is the number of active-space spin orbitals. To efficiently suppress this sampling burden on near-term quantum hardware, several advanced measurement reduction and state tomography strategies can be seamlessly integrated into our workflow.

 A primary strategy to reduce the measurement overhead is to exploit Pauli grouping techniques. These methods leverage the fact that mutually commuting observables share a common set of eigenstates and can therefore be measured simultaneously. The fermionic operators corresponding to the $k$-RDM are first mapped to linear combinations of Pauli strings, which can then be grouped based on two types of commutation relations: qubit-wise commutativity (QWC) and general commutativity (GC). QWC requires that for each qubit, the Pauli operators in the strings commute individually, while GC requires only that the full Pauli strings commute as operators. Specifically, two Pauli strings commute under GC if their Pauli operators anti-commute on an even number (including zero) of qubit positions. As a result, QWC is a subset of GC and is computationally simpler to implement. Graph-based algorithms are often used to identify commuting groups under both QWC and GC frameworks~\cite{tilly2022variational,verteletskyi2020measurement,gokhale2019minimizing}.

Furthermore, the classical shadow~\cite{huang2020predicting} method offers an alternative paradigm. It employs randomized measurements to approximate quantum states and compute expectation values of observables, enabling partial tomography. Zhao et al. proposed a fermionic classical shadow protocol based on fermionic Gaussian unitaries (FGU)~\cite{zhao2021fermionic}. Unlike standard Pauli-based shadows where sample complexity scales exponentially with the weight of the mapped Pauli strings (which is problematic for non-local fermionic operators), the FGU-based shadow is tailored for fermionic systems. According to Zhao et al., the number of measurements ($M$) required to estimate all elements of a fermionic $k$-RDM within an additive error $\epsilon$ scales as:
$$ M = O\left( \binom{N_{\rm act}}{k} k^{3/2} \log(N_{\rm act}) / \epsilon^2 \right) $$
For both VQE-PT and VQE-PTs, which require up to 4-RDMs ($k=4$), the measurement cost under this protocol scales as $\tilde{O}(N_{\rm act}^4)$. This confirms that extracting the necessary high-order RDMs is polynomial in the active-space size, in contrast to the exponential scaling of full state tomography.


The cumulant expansion method~\cite{harris2002cumulant, takemori2023balancing} provides an exact decomposition of a $k$-RDM into connected RDMs ($^k\Delta$) using the wedge product $\wedge$, which denotes the antisymmetrized tensor product. For instance, the first few orders are
\begin{equation}
\begin{split}
   ^1D&=^{1}\Delta\\ 
   ^2D&=^2\Delta + ^1\Delta\wedge^1\Delta\\
   ^3D&=^3\Delta + 3^2\Delta\wedge^1\Delta + ^1\Delta\wedge^1\Delta\wedge^1\Delta
\end{split}
\end{equation}
In practice, the cumulant approach can approximate high-order RDMs by truncating the expansion—neglecting higher-order connected RDMs and retaining only lower-order terms.
For example, if only the 1-RDM and 2-RDM are available, the 3-RDM can be approximated as
\begin{equation}
^3D \approx 3^2\Delta\wedge^1\Delta + ^1\Delta\wedge^1\Delta\wedge^1\Delta .
\end{equation}
This approximation is particularly advantageous when the number of samples is limited or measurement noise is significant.

\section{Spin-Adapted Excitation Operators}\label{sec:adapt}
The spin-adapted excitation operators are defined as~\cite{helgaker2013molecular}
 \begin{equation}
\begin{split}
&\hat{E}^{p}_q =\hat{a}_p^\dagger \hat{a}_q + \hat{a}^\dagger_{\overline{p}}\hat{a}_{\overline{q}}\\
&\hat{E}^{pq,(+)}_{rs}
=
\hat{E}^{p}_{r}\hat{E}^{q}_{s}
+
\hat{E}^{p}_{s}\hat{E}^{q}_{r},\\
&\hat{E}^{pq,(-)}_{rs}
=
\hat{E}^{p}_{r}\hat{E}^{q}_{s}
-
\hat{E}^{p}_{s}\hat{E}^{q}_{r}.
\end{split}
\end{equation}
 Here, a bar over an index denotes a spin-down orbital. 
 \section{Effect of Basis-Set Expansion}

\begin{figure}[H]   
	\centering	\includegraphics[width=1\linewidth]{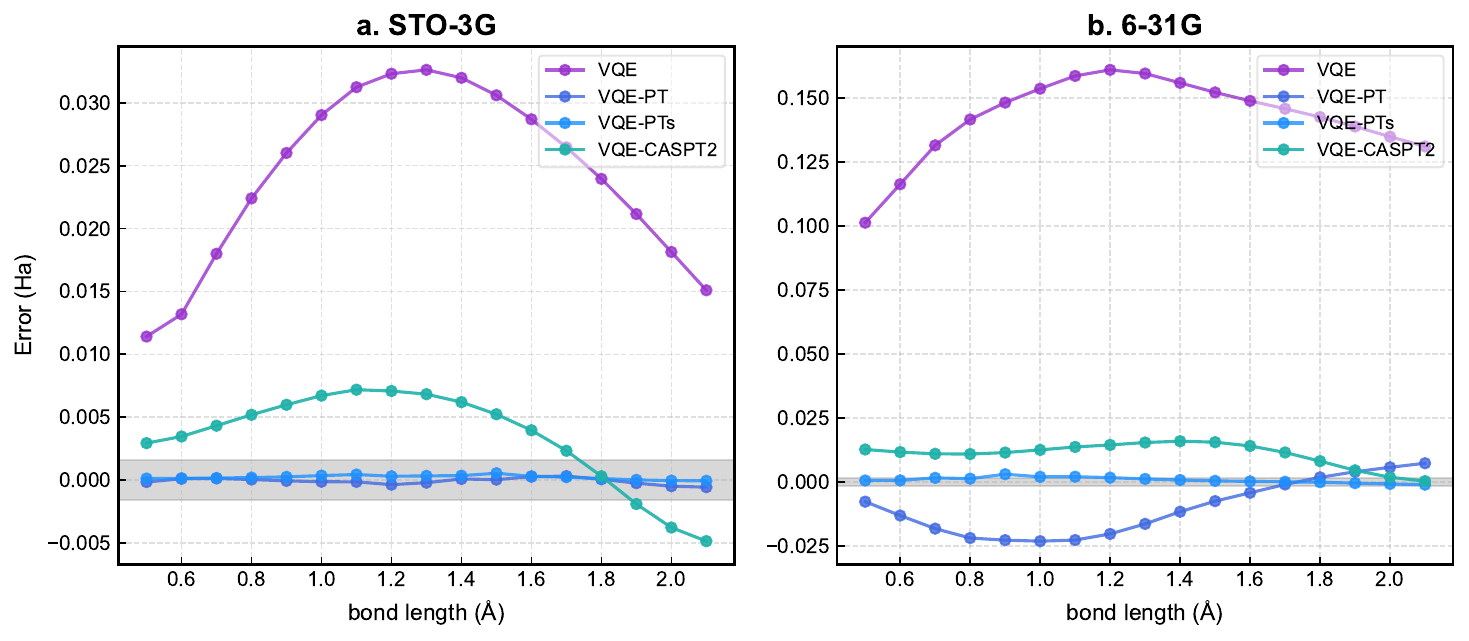}
	\caption{{Effect of basis-set expansion on perturbative corrections for the ground state of the N$_2$ molecule. Energy deviations relative to the corresponding FCI reference are shown for active-space VQE, VQE-PT, VQE-PTs, and VQE-CASPT2 with (a) STO-3G and (b) 6-31G basis sets using a fixed (6e, 6o) active space. The shaded region indicates the chemical accuracy threshold.}}
	\label{fig:large_basis}
\end{figure}

Having established the accuracy of the perturbative VQE framework in small basis sets in the main text, we examine the effect of basis-set expansion. We expand the basis set of \ce{N2} from STO-3G to 6-31G while keeping the (6e, 6o) active space fixed. Fig.~\ref{fig:large_basis} shows the energy deviations of active-space VQE, VQE-PTs, VQE-PT, and VQE-CASPT2 from the corresponding FCI reference energies.

\begin{table}[!htbp]
\caption{Mean absolute errors (MAE, in mHa) of ground-state energies for active-space VQE, VQE-PT, VQE-PTs, and VQE-CASPT2, relative to the corresponding FCI benchmarks for different basis sets.}
\label{table:mae_basis_active}
\centering
\small
\begin{tabular}{ccccc}
       \toprule
       \midrule
        Setting & VQE & VQE-PT & VQE-PTs & VQE-CASPT2 \\
       \midrule
        STO-3G & 24.25 & 0.20 & 0.23 & 4.60 \\     
        6-31G  & 142.41 & 12.32 & 1.05 & 10.90 \\

       \bottomrule
\end{tabular}
\end{table}

When the basis set is enlarged from STO-3G to 6-31G while the active space is fixed at (6e, 6o), the mean absolute error (MAE) of the active-space VQE energy increases from 24.25 to 142.41 mHa. This should be interpreted as an active-space truncation effect rather than a loss of accuracy of the VQE solver within the selected active space. By incorporating external correlation through perturbative corrections, both VQE-PT and VQE-PTs substantially reduce the active-space truncation error. In particular, VQE-PT lowers the MAE to 12.32 mHa, recovering most of the missing correlation effects. The fully coupled VQE-PTs treatment further improves the accuracy, reducing the MAE to 1.05 mHa, which corresponds to a reduction in error of more than 99\% relative to the uncorrected active-space VQE reference. The larger improvement of VQE-PTs over VQE-PT for the 6-31G basis set suggests that explicitly accounting for the couplings among internally contracted perturbers becomes increasingly beneficial when the external correlation space grows. For comparison, VQE-CASPT2 also provides a significant improvement over the active-space VQE reference, achieving an MAE of 10.90 mHa, but remains less accurate than the fully coupled VQE-PTs approach.

\section{Evaluation of Active Space Size}
\begin{figure}[H]   
	\centering	\includegraphics[width=0.8\linewidth]{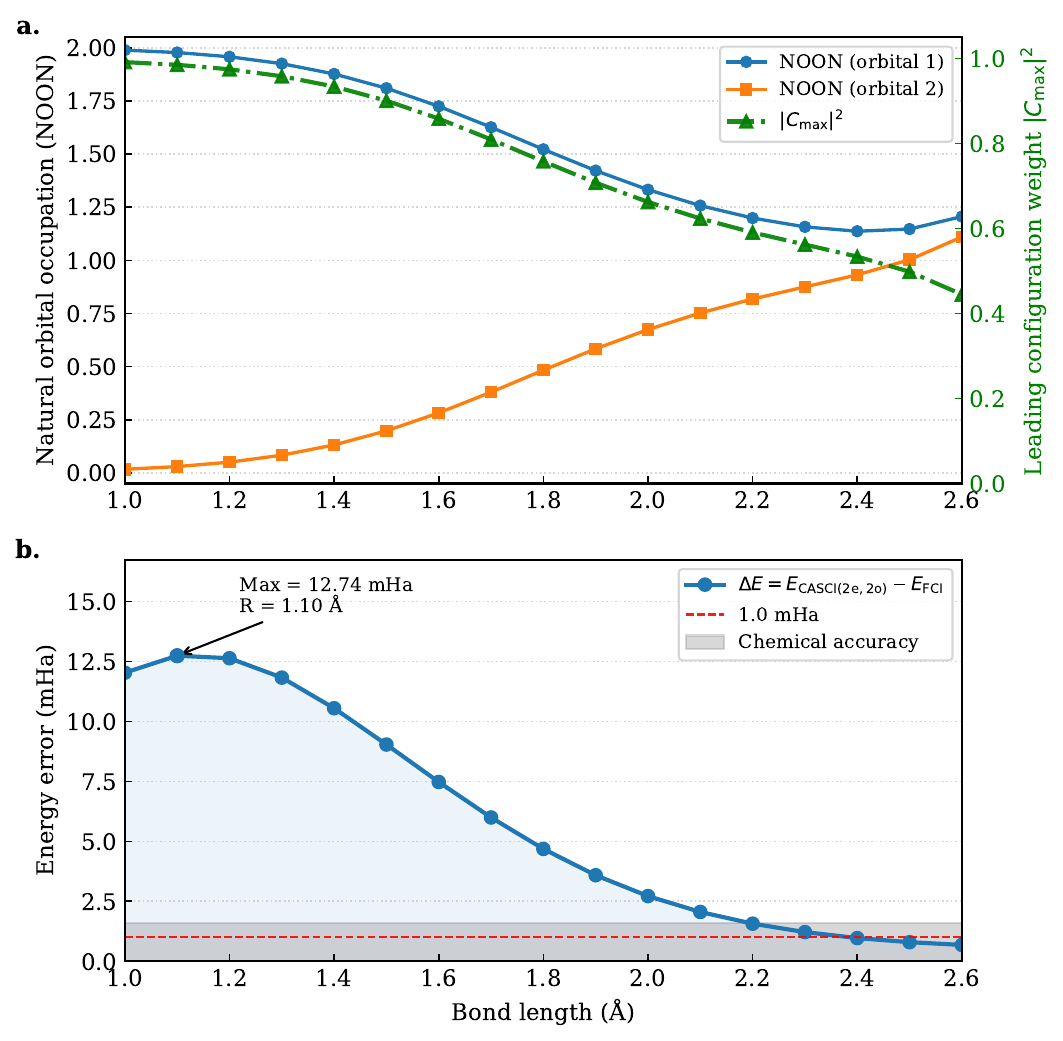}
	\caption{(a) Active-space natural orbital occupation numbers (NOONs) computed from CASCI(2e, 2o) (left axis) and the weight of the leading configuration $|C_\text{max}|^2$ obtained from FCI (right axis) for F$_2$ molecule as a function of bond length. The calculation is performed in the STO-6G basis set. The convergence of NOONs to $\approx 1.0$ and the decay of $|C_\text{max}|^2$ indicate a transition to a strongly statically correlated regime. (b) Energy difference between CASCI(2e, 2o) and FCI calculations for F$_2$. The error is largest near equilibrium where dynamic correlation dominates, and minimizes at the dissociation limit where static correlation dominates. This confirms that the (2e, 2o) active space is sufficient to capture the static correlation effects.}
	\label{fig:NOON}
\end{figure}

To assess the accuracy of selecting the (2e, 2o) active space for the zeroth-order description of the ground-state of F$_2$ (STO-6G) in our experiment, we correlated wavefunction diagnostics (CASCI natural orbital occupations (NOONs) and FCI leading configuration weight $|C_\text{max}|^2$) with the energy error relative to the FCI (whole space (18e, 10o)) benchmark, shown in Fig.~\ref{fig:NOON}. Near equilibrium, the system exhibits single-reference character ($|C_\text{max}|^2 \approx 0.99$, NOONs $\approx$ 2.0/0.0), yet the energy deviation from FCI is maximized ($\approx 12$ mHa). This indicates that the error in this region is predominantly dynamic correlation, which is missing in the small CAS space. Conversely, towards dissociation, strong static correlation emerges as $|C_\text{max}|^2$ drops to $\approx 0.44$ and NOONs approach (1.06, 0.94). Crucially, the energy error simultaneously drops below 1 mHa. This contrast demonstrates that the minimal (2e, 2o) space successfully encapsulates the dominant static correlation required for bond breaking, leaving the dynamic correlation (dominant at equilibrium) to be recovered by the subsequent VQE-PT treatment.

\bibliography{qc}

\end{document}